\documentclass[preprint,3p]{elsarticle} 

\usepackage{graphicx}
\usepackage{algorithm}
\usepackage{algorithmic}
\usepackage{parskip}
\usepackage{color}
\usepackage{tikz}
\usepackage{multirow}

\newtheorem{definition}{Definition}
\newtheorem{theorem}{Theorem}[section]

\usepackage{amssymb,amsmath}

\journal{Knowledge Based Systems}

\begin{document}

\begin{frontmatter}



\title{A Novel Measure of Edge Centrality in Social Networks}


\author[1]{Pasquale De Meo}
\address[1]{Department of Physics, Informatics Section}
\ead{pdemeo@unime.it}
\author[2]{Emilio Ferrara\corref{cor1}}
\ead{eferrara@unime.it}
\cortext[cor1]{Corresponding author}
\author[1]{Giacomo Fiumara}
\ead{gfiumara@unime.it}
\author[2]{Angela Ricciardello}
\address[2]{Department of Mathematics\\University of Messina, V.le F. Stagno D'Alcontres 31, 98166 (ME), Italy}
\ead{aricciardello@unime.it}

\begin{abstract}{} 
The problem of assigning centrality values to nodes and edges in graphs has been widely
investigated during last years. Recently, a novel measure of node centrality has been proposed,
called $\kappa$-path centrality index, which is based on the propagation of messages inside a
network along paths consisting of at most $\kappa$ edges. On the other hand, the importance of
computing the centrality of edges has been put into evidence since 1970's by Anthonisse and,
subsequently by Girvan and Newman. In this work we propose the generalization of the concept of
$\kappa$-path centrality by defining the $\kappa$-path edge centrality, a measure of
centrality introduced to compute the importance of edges. We provide an efficient algorithm,
running in $O(\kappa m)$, being $m$ the number of edges in the graph. Thus, our technique is
feasible for large scale network analysis. Finally, the performance of our algorithm is analyzed,
discussing the results obtained against large online social network datasets.
\end{abstract}


\end{frontmatter}

\section{Introduction} \label{sec:introduction}
In the context of the social knowledge management, Social Network Analysis (SNA) is attracting an
increasing attention by the scientific community, in particular during the latest years. 
One of the main motivations is the unprecedented success of phenomena such as online social networks and online
communities. In this panorama, not only from a scientific perspective but also for commercial or
strategic motivations, the identification of the principal actors inside a network is very
important.

Such an identification requires to define an {\em importance measure} (also referred to as {\em
centrality}) to weight nodes and/or edges.

The simplest approaches to computing centrality consider only the {\em local topological
properties} of a node/edge in the social network graph: for instance, the most intuitive node
centrality measure is represented by the degree of a node, i.e., the number of social contacts of a
user. Unfortunately, local measures of centrality, whose esteem is computationally feasible even on
large networks, do not produce very faithful results \cite{Carrington2005}.

Due to these reasons, many authors suggested to consider the {\em whole social network topology} to
compute centrality values. A new family of centrality measures was born, called {\em global
measures}. Some examples of global centrality measures are {\em closeness}
\cite{sabidussi1966centrality} and {\em betweenness centrality} (for nodes \cite{freeman1977set},
and edges \cite{anthonisse1971rush,girvan2002community}).

Betweenness centrality is one of the most popular measures and its computation is the core component
of a range of algorithms and applications.
Betweenness centrality relies on the idea that, in social
networks, information flows along {\em shortest paths}: as a consequence, a node/edge has a high
betweenness centrality if a large number of shortest paths crosses it.

Some authors, however, raised some concerns on the effectiveness of betweenness centrality. First
of all, the problem of computing the exact value of betweenness centrality for each node/edge of a
given graph is computationally demanding -- or even unfeasible -- as the size of the analyzed
network grows. Therefore, the need of finding fast, even if approximate, techniques to compute
betweenness centrality arises and it is currently a relevant research topic in Social Network
Analysis.

A further issue is that the assumption that information in social networks propagates only along
shortest paths could not be true \cite{stephenson1989rethinking}. By contrast, information
propagation models have been provided in which information, encoded as messages generated in a
source node and directed toward a target node in the network, may flow along {\em arbitrary} paths.
In the spirit of such a model, some authors \cite{newman2005measure,noh2004random} suggested to
perform random walks on the social network to compute centrality values.

A prominent approach following this research line is the work proposed in
\cite{alahakoon2011kpath}. In that work, the authors introduced a novel node centrality measure
known as $\kappa$-{\em path centrality}. In detail, the authors suggested to use self-avoiding
random walks \cite{madras1996self} of length $\kappa$ (being $\kappa$ a suitable integer) to
compute centrality values. They provided an approximate algorithm, running in $O(\kappa^3 \
n^{2-2\alpha} \log \ n)$ being $n$ the number of nodes and $\alpha \in [-\frac{1}{2},\frac{1}{2}]$.

In this paper we extend that work \cite{alahakoon2011kpath} by introducing a measure of {\em
edge centrality}. This measure is called $\kappa$-{\em path edge centrality}. In our approach, the
procedure of computing edge centrality is viewed as an {\em information propagation problem}. In
detail, if we assume that multiple messages are generated and propagated within a social network,
an edge is considered as ``central'' if it is frequently exploited to diffuse information.

Relying on this idea, we simulate message propagations through random walks on the social network
graphs. In our simulation, in addition, we assume that random walks are {\em simple} and of {\em
bounded length} up to a constant and user-defined value $\kappa$. The former assumption is because
a random walk should be forced to pass no more than once through an edge; the latter, because, as
in \cite{friedkin1983horizons}, we assume that the more distant two nodes are, the less they
influence each other.

The computation of edge centrality has many practical applications in a wide range of contexts and,
in particular, in the area of Knowledge-Based (KB) systems. For instance in KB systems in which
data can be conveniently managed through graphs, the procedure of weighting edges plays a key role
in identifying {\em communities}, i.e., groups of nodes densely connected to each other and weakly
coupled with nodes residing outside the community itself \cite{SrMu94,xia2011community}. This is
useful to better organize available knowledge: think, for instance, to an e-commerce platform and
observe that we could partition customer communities into smaller groups and we could selectively
forward messages (like commercial advertisements) only to groups whose members are actually
interested to them. In addition, in the context of Semantic Web, edge centralities are useful to
quantify the strength of the relationships linking two objects and, therefore, it can be useful to
discover new knowledge \cite{RoWa10}. Finally, in the context of social networks, edge centralities
are helpful to model the intensity of the social tie between two individuals
\cite{Ding*11}: in such a case, we could extract patterns of interactions among users in
virtual communities and analyze them to understand how a user is able to influence another one.
The main contributions of this paper are the following:

\begin{itemize}
\item We propose an approach based on {\em random walks} consisting of up-to $\kappa$ edges to
    compute {\em edge centrality}. In detail, we observe that many approaches in the literature
    have been proposed to compute node centrality but, comparatively, there are few studies on
    edge centrality computation (among them we cite the edge betweenness centrality introduced
    in the Girvan-Newman algorithm \cite{girvan2002community}). In addition, some authors
    \cite{newman2005measure,noh2004random,brandes2005centrality} successfully applied random
    walks to compute node centrality in networks. We suggest to extend these ideas in the
    direction of edge centrality, and, therefore, this work is the {\em first attempt} to
    compute edge centrality by means of random walks.

\item We design an algorithm to efficiently compute edge centrality.
	 	The worst case time complexity of our algorithm is $O(\kappa m)$, being $m$ the number of edges in the social network graph and $\kappa$ a constant (and typically small) factor.
		Therefore, the running time of our algorithm scales in {\em linear fashion} against the number
		of edges of a social network. This is an interesting improvement of the state-of-the-art: in fact, {\em
		exact algorithms} for computing centrality run in $O(n^3)$
		and, with some ingenious optimizations they can run in $O(nm)$
		\cite{brandes2001faster,girvan2002community}. Unfortunately, real-life social networks consist
		of up to millions nodes/edges \cite{mislove2007measurement}, and, therefore these approaches may
		not scale well. By contrast, our algorithm works fairly well also on large real-life social
		networks even in presence of limited computing resources.

\item We provide results of the performed experimentation, showing that our approach is able to generate
    reproducible results even if it relies on random walks.
		Several experiments have been carried out in order to emphasize that the
		$\kappa$-path edge centrality computation is feasible even on large social networks.
		Finally, the properties shown by this measure are discussed, in order to
		characterize each of the studied networks.
\end{itemize}

The paper is organized as follows: in section \ref{sec:background} we provide some background
information on the problems related to centrality measures. Section \ref{measuring-edge-centrality}
presents the goal of this paper and our $\kappa$-path edge centrality, including the fast
algorithm for its computation. The experimental evaluation of performance of this strategy
is discussed in section \ref{sec:experimentation} and some possible applications of our approach
are presented in section \ref{sec:applications}. Thus, the paper concludes in section
\ref{sec:conclusions}.

\section{Background about Centrality Measures and Applications} \label{sec:background}

In this section we review the concept of centrality measure and illustrate some recent approaches to compute it.

\subsection{Centrality Measure in Social Networks} \label{sub:centralitydef}

One of the first (and the most popular) node centrality measures is the \emph{betweenness
centrality} \cite{freeman1977set}. It is defined as follows:

\begin{definition}(Betweenness centrality)
Given a graph $G = \langle V, E \rangle$, the betweenness centrality for the node $v \in V$ is defined as

\begin{equation}
C_{B_n}(v)= \sum_{s \neq v \neq t \in V}{\frac{\sigma_{st}(v)}{\sigma_{st}}}
\label{eq:betweenness-centrality}
\end{equation}

where $s$ and $t$ are nodes in $V$, $\sigma_{st}$ is the number of shortest paths connecting $s$ to
$t$, and $\sigma_{st}(v)$ is the number of shortest paths connecting $s$ to $t$ passing through the
node $v$.
\end{definition}

If there is no path joining $s$ and $t$ we conventionally set $\frac{\sigma_{st}(v)}{\sigma_{st}} =
0$.

The concept of centrality has been defined also for the edges in a graph and, from a historical
standpoint, the first approach to compute edge centrality has been proposed in 1971 by J.M. Anthonisse
\cite{anthonisse1971rush,brandes2005introduction} and was implemented in the GRADAP software
package. In this approach, edge centrality is interpreted as a ``flow centrality'' measure. To
define it, let us consider a graph $G = \langle V, E \rangle$ and let $s \in V$, $t \in V$ be a
{\em fixed} pair of nodes. Assume that a ``unit of flow'' is injected in the network by picking $s$
as the source node and assume that this unit flows in $G$ along the shortest paths. The {\em rush
index} associated with the pair $\langle s, t \rangle$ and the edge $e \in E$ is defined as

$$
\delta_{st}(e) = \frac{\sigma_{st}(e)}{\sigma_{st}}
$$

being, as before, $\sigma_{st}$ the number of shortest paths connecting $s$ to $t$, and
$\sigma_{st}(e)$ the number of shortest paths connecting $s$ to $t$ passing through the edge $e$.
As in the previous case, we conventionally set $\delta_{st}(e) = 0$ if there is no path joining $s$
and $t$.

The rush index of an edge $e$ ranges from 0 (if $e$ {\em does not belong} to any shortest path
joining $s$ and $t$) to 1 (if $e$ belongs to {\em all} the shortest paths joining $s$ and $t$).
Therefore, the higher $\delta_{st}$, the more relevant the contribution of $e$ in the transfer of a
unit of flow from $s$ to $t$. The {\em centrality} of $e$ can be defined by considering all the
pairs $\langle s,t \rangle$ of nodes and by computing, for each pair, the rush index
$\delta_{st}(e)$; the centrality $C_{R_e}(e)$ of $e$ is the sum of all these contributions

$$
C_{R_e}(e) = \sum_{s \in V} \sum_{v \in V} \delta_{st}(e)
$$

More recently, in 2002 Girvan and Newman proposed in \cite{girvan2002community} a definition of
\emph{edge betweenness centrality} which strongly resembles that provided by Anthonisse.

According to the notation introduced above, the edge betweenness centrality for the edge $e \in E$
is defined as

\begin{equation}
C_{B_e}(e) = \sum_{s \neq t \in V}{ \frac{\sigma_{st}(e)}{\sigma_{st}}}
\label{eq:edge-betweenness-centrality}
\end{equation}

and it differs from that of Anthonisse because the source node $s$ and the target node $t$ must be
different.

Other, marginally different, definitions of betweenness centrality have been proposed by
\cite{brandes2008variants}, such as bounded-distance, distance-scaled, edge and group betweenness,
and stress and load centrality.

Although the appropriateness of the betweenness centrality in the representation of the
``importance'' of a node/edge inside the network is evident, its adoption is not always the unique
solution to a given problem. For example, as already put into evidence by
\cite{stephenson1989rethinking}, the first limit of the concept of betweenness centrality is
related to the fact that influence or information does not propagate following only shortest paths.
With regards to the influence propagation, it is also evident that the more distant two nodes are,
the less they influence each other, as stated by \cite{friedkin1983horizons}. Additionally, in real
applications (such as those described in section \ref{sec:application}) it is not usually required
to calculate the exact ranking with respect to the betweenness centrality of each node/edge inside
the network. In fact, it results more useful to identify the top arbitrary percentage of
nodes/edges which are more relevant to the given specific problem (e.g., study of propagation of
information, identification of key actors, etc.).

\subsection{Recent Approaches for Computing Betweenness Centrality} \label{sec:fast-betweenness}

As to date, several algorithms  to compute the betweenness centrality (of nodes) in a graph have
been presented. The most efficient has been proposed by  \cite{brandes2001faster}, which runs in
$O(n m)$ for unweighted graphs, and in $O(n m + n^2 log \ n)$ for weighted graphs, containing $n$
nodes and $m$ edges.

The computational complexity of these approaches makes them unfeasible for large network analysis.
To this purpose, different approximate solutions have been proposed. Amongst others,
\cite{brandes2007centrality} developed a randomized algorithm (namely, ``RA-Brandes'') and,
similarly by using adaptive techniques, \cite{bader2007approximating} proposed another approximate
version (called, ``AS-Bader''). In \cite{newman2005measure}, Newman devised a random-walk based
algorithm to compute betweenness centrality which shares similarities to our approach, starting
from the concept of message propagation along random paths. From the same concept,
\cite{alahakoon2011kpath} proposed the $\kappa$-path centrality measure (for nodes) and developed a
$O(\kappa^3 \ n^{2-2\alpha} \log \ n)$ algorithm (namely, ``RA-$\kappa$path'') to compute it.

\subsection{Application of Centrality Measures in Social Network Analysis} \label{sec:application}
Applications of centrality information acquired from social networks have been investigated by
\cite{Staab2005}. The authors defined different methodologies to exploit discovered data, e.g., for
marketing purposes, recommendation and trust analysis.

Several marketing and commercial studies have been applied to online social networks
(OSNs), in particular to discover efficient channels to distribute information
\cite{brown2007word,trusov2009effects} and to study the spread of influence \cite{kempe2003maximizing}.
Potentially, our study could provide useful information to all these applied research directions, identifying those
interesting edges with high $\kappa$-path edge centrality, which emphasizes their importance within
the social network.
Those nodes interconnected by high central edges are important because of the
position they ``topologically'' occupy. Moreover, they could efficiently carry information to their
neighborhood.

\section{Measuring Edge Centrality} \label{measuring-edge-centrality}

\subsection{Design Goals} \label{sec:sketch}
Before to providing a formal description of our algorithm, we illustrate the main ideas behind it.
We start from a real-life example and we use it to derive some ``requirements'' our algorithm should satisfy.

Let us consider a network of devices. In this context, without loss of generality, we
can assume that the simplest ``piece'' of information is a message. In addition, each device has an
{\em address book} storing the devices with which it can exchange messages. A device can both
receive and transmit messages to other devices appearing in its address book.

The purpose of our algorithm is to rank links of the network on the basis of their aptitude of favoring the diffusion of information.
In detail, the higher the rank of a link, the higher its ability of propagating a message.
Henceforth, we refer to this problem as \emph{link ranking}.

The link ranking problem in our scenario can be viewed as the problem of computing edge centrality
in social networks. We guess that some of the hypotheses/procedures adopted to compute edge centrality
can be applied to solve the link ranking problem. We suggest to extend these techniques in a number of ways.
In detail, we guess that the algorithm to compute the link ranking should satisfy the following
requirements:

{\em Requirement 1 - Simulation of Message Propagation by using Random Walks}.
	As shown in section \ref{sec:background}, some authors assume that information flows on a network along the shortest paths.
	Such an intuition is formally captured by Equation	(\ref{eq:betweenness-centrality}).
	However, as observed in \cite{freeman1991cen,newman2005measure}, centrality measures based on shortest paths can provide some counterintuitive results. In detail, \cite{freeman1991cen,newman2005measure} present some simple examples
	showing that the application of Equation (\ref{eq:betweenness-centrality}) would lead to
	assign excessively low centrality scores to some nodes.

	To this purpose, some authors \cite{freeman1991cen} provided a more refined definition of
	centrality relying on the concept of {\em flow} in a graph. To define this measure, assume that
	each edge in the network can carry one or more messages; we are interested in finding those
edges
	capable of transferring the largest amount of messages between a source node $s$ and a target
	node $t$. The centrality of a vertex $v$ can be computed by considering all the pairs $\langle
	s,t \rangle$ of nodes and, for each pair, by computing the amount of flow passing through $v$. In
	the light of such a definition, in the computation of node centrality also non-shortest paths
	are considered.

	However, in \cite{newman2005measure}, Newman shows that centrality measures based on the concept of flow are not exempt from odd effects.
	To this purpose, the author suggests to consider a random walker which \emph{is not forced} to
	move along the shortest paths of a network to compute the centrality of nodes.

	The Newman's strategy has been designed to compute node centrality, whereas our approach targets at computing edge centrality.
 	Despite this difference, we believe that the idea of using random walks in place of shortest paths can be successful even when applied to the link ranking problem.

   In our scenario, if a device wants to propagate a message, it is generally not aware of the
   whole network topology, and therefore it is not aware of the shortest paths to route the
   message. In fact, each device is only aware of the devices appearing in its address book.
   As a consequence, the device selects, according to its own criteria, one (or more) of its
   contacts and sends them the message in the hope that they will further continue the
   propagation. In order to simulate the message propagation, our first requirement is to
   exploit random walks.

{\em Requirement 2 - Dynamic Update of Ranking}.
	Ideally, if we would simulate the propagation of multiple messages on our network of devices, it could happen that an edge is selected more frequently than others.
	Edges appearing more frequently than others show a better aptitude to spread messages and,
    therefore, their rank should be higher than others.
	As a consequence, our mechanism to rank edges should be {\em dynamic}: at the beginning, all the edges are equally likely to propagate a message and, therefore, they have the same rank.
	At each step of the simulation, if an edge is selected, it must be awarded by getting a ``bonus
score''. 	

{\em Requirement 3 - Simple Paths}.
	The procedure of simulating message propagation through random walks described above could
imply that a message can pass through an edge more than once. In such a case, the rank of edges
which are traversed multiple times would be disproportionately inflated whereas the rank of edges
rarely (or never) visited could be underestimated. The global effect would be that the ranking
produced by this approach would not be correct.	As a consequence, another requirement is that the
paths exploited by our algorithm must be simple.

{\em Requirement 4 - Bounded Length Paths}.
	As shown in \cite{friedkin1983horizons}, the more distant two nodes are, the less they
influence each other.
	The usage of paths of bounded length has been already explored to compute node centrality
\cite{borgatti2006graph,everett2005ego}. A first relevant example is provided in
\cite{everett2005ego}; in that paper the authors observe that methods to compute node centralities
like those based on eigenvectors can lead to counterintuitive results.
	In fact, those methods take the whole network topology into account and, therefore, they
compute the centrality of a node on a {\em global scale}.
	It may happen that a node could have a big impact on a small scale (think of a well-respected
researcher working on a niche topic) but a limited visibility on a large scale.
	Therefore, the approach of \cite{everett2005ego} suggested to compute node centralities in local networks and they considered {\em ego networks}.
	An ego network is defined as a network consisting of a single node ({\em ego}) together with the nodes it is connected to (the {\em alters}) and all the links among those alters.
	The diameter of an ego network is 2 and, therefore, the computation of node centrality in a network requires to compute paths up to a length 2.
	In \cite{borgatti2006graph} the authors extended these concepts by considering paths up to a length $k$.

	We agree with the observations above and figure that two nodes are considered to be distant if the shortest path connecting them is longer than $\kappa$ hops, being $\kappa$ the established threshold.
	Such a consideration depicts as effective paths only those paths whose length is up to $\kappa$.
	We take this requirement and for our simulation procedure we considered paths of bounded length.

	In the next sections we shall discuss how our algorithm is able to incorporate the requirements illustrated above.

\subsection{{\LARGE$\kappa$}-Path Centrality} \label{sub:kpathcentrality}

In this section we introduce the concepts of $\kappa$-path node centrality and $\kappa$-path edge centrality.

The notion of $\kappa$-path node centrality, introduced by \cite{alahakoon2011kpath}, is defined as follows:

\begin{definition} ($\kappa$-path node centrality) For each node $v$ of a graph \mbox{$G = \langle V,E \rangle$}, the $\kappa$-path node centrality $C^\kappa(v)$ of $v$ is defined as the sum, over all possible source nodes $s$, of the frequency with which a message originated from $s$ goes through $v$, assuming that the message traversals are only along random simple paths of at most $\kappa$ edges.
\label{def:k-path-centrality}
\end{definition}

It can be formalized, for an arbitrary node $v \in V$, as

\begin{equation}
C^\kappa(v) = \sum_{s \in V}{\frac{\sigma_s^\kappa(v)}{\sigma_s^\kappa}}
\label{eq:k-path}
\end{equation}
where $s$ are all the possible source nodes, $\sigma_s^\kappa(v)$ is the number of $\kappa$-paths originating from $s$ and passing through $v$ and $\sigma_s^\kappa$ is the overall number of $\kappa$-paths originating from $s$.

Observe that Equation (\ref{eq:k-path}) resembles the definition of \emph{betweenness centrality} provided in Equation (\ref{eq:betweenness-centrality}).
In fact, the structure of the two equations coincides if we replace the concept of shortest paths (adopted in the betweenness centrality) with the concept of $\kappa$-paths which is the core of our definition of $\kappa$-path centrality.

The possibility of extending the concept of ``centrality'' from nodes to edges has been already exploited by Girvan and Newman \cite{girvan2002community}.
In particular, they generalized the formulation of ``betweenness centrality'' (referred to nodes), introducing the concept of ``edge betweenness centrality''.

Similarly, we extend Definition \ref{def:k-path-centrality} in order to define an edge centrality index, baptized \emph{$\kappa$-path edge centrality}.

\begin{definition} ($\kappa$-path edge centrality) For each edge $e$ of a graph \mbox{$G = \langle V,E \rangle$},
the $\kappa$-path edge centrality $L^\kappa(e)$ of $e$ is defined as the sum, over all possible
source nodes $s$, of the frequency with which a message originated from $s$ traverses $e$, assuming
that the message traversals are only along random simple paths of at most $\kappa$ edges.
\label{def:k-path-edge-centrality}
\end{definition}

The \emph{$\kappa$-path edge centrality} is formalized, for an arbitrary edge \emph{e}, as follows
\begin{equation}
L^\kappa(e) = \sum_{s \in V}{\frac{\sigma_s^\kappa(e)}{\sigma_s^\kappa}}
\label{eq:edge-k-path}
\end{equation}

where $s$ are all the possible source nodes, $\sigma_s^\kappa(e)$ is the number of $\kappa$-paths
originating from $s$ and traversing the edge $e$ and, finally, $\sigma_s^\kappa$ is the number of
$\kappa$-paths originating from $s$.

In practical cases, the application of Equation (\ref{eq:edge-k-path}) can not be feasible because
it requires to count all the $\kappa$-paths originating from all the source nodes $s$ and such a number can
be exponential in the number of nodes of $G$. To this purpose, we need to design some algorithms capable
of efficiently approximating the value of $\kappa$-path edge centrality. These algorithms will be
introduced and discussed in the next subsections.

\subsection{The Algorithm for Computing the  {\LARGE $\kappa$}-Path Edge Centrality} \label{sec:edge-k-path-centrality}

In this section we discuss an algorithm, called \emph{Edge Random Walk $\kappa$-Path Centrality}
(or, shortly, {\em ERW-Kpath}), to efficiently compute edge centrality values.

It consists of two main steps: \emph{(i)} node and edge weights assignment and, \emph{(ii)}
simulation of message propagations through random simple paths. In the ERW-KPath algorithm, the
probability of selecting a node or an edge are uniform; we provide also another version of the
ERW-Kpath algorithm (called {\em WERW-Kpath - Weighted Edge Random Walk $\kappa$-Path Centrality})
in which the node/edge probabilities are not uniform.

We will show in the Appendix that the ERW-KPath and the WERW-Kpath algorithms return, as output, an
approximate value of the edge centrality index as provided in Definition
\ref{def:k-path-edge-centrality} and we will provide a quantitative assessment of such an
approximation.

In the following we shall discuss the ERW-KPath algorithm by illustrating each of the two steps composing it.
After that, we will introduce the WERW-KPath algorithm as a generalization of the ERW-KPath algorithm.

\subsubsection{Step 1: node and edge weights assignment}

In the first stage of our algorithm, we assign a weight to both nodes and edges of the graph $G =
\langle V,E \rangle $ representing our social network. Weights on nodes are used to select the
source nodes from which each message propagation simulation starts. Weights on edges represent initial values of
edge centrality and, to comply with Requirement 2, they will be updated during the execution of our
algorithm.

To compute weight on nodes, we introduce the \emph{normalized degree} $\delta(v_n)$ of a node $v_n
\in V$ as follows:

\begin{definition}{(Normalized degree)}
Given an undirected graph $G=\langle V,E \rangle$ and a node $v_n \in V$, its normalized degree
$\delta(v_n)$ is
\begin{equation}
	\delta(v_n) = \frac{|I(v_n)|}{|V|}
	\label{eq:local-density}
\end{equation}
where $I(v_n)$ represents the set of edges incident on $v_n$.
\end{definition}

The normalized degree $\delta(v_n)$ correlates the degree of $v_n$ and the number of total nodes on the network.
Intuitively, it represents how much a node contributes to the overall connectivity of the graph.
Its value belongs to the interval $[0,1]$ and the higher $\delta(v_n)$, the better $v_n$ is connected in the graph.

Regarding edge weights, we introduce the following definition:

\begin{definition}{(Initial edge weight)}
Given an undirected graph $G=\langle V,E \rangle$ and an edge $e_m \in E$, its initial edge weight
$\omega_0(e_m)$ is
\begin{equation}
	\omega_0(e_m)  = \frac{1}{|E|}
\label{eq:initial-edge-weight}
\end{equation}
\end{definition}

Intuitively, the meaning of Equation (\ref{eq:initial-edge-weight}) is as follows: we initially
manage a ``budget'' consisting of $|E|$ points; these points are equally divided among all the possible
edges; the amount of points received by an edge represents its initial rank.

\begin{figure}[!ht]%
	\centering 	
	\tikzstyle{node}=[circle,fill=black!25,minimum size=20pt,inner sep=0pt]
	\tikzstyle{edge} = [draw,thick,-]
	\tikzstyle{weight} = [font=\footnotesize] 	
	\begin{tikzpicture}
  [scale=.8,auto=left]
  \node (a) at (-2,-2) {a({\color{blue}$\frac{1}{11}$})};
  \node (b) at (1,0)  {b({\color{blue}$\frac{3}{11}$})};
  \node (c) at (4,-2) {c({\color{blue}$\frac{2}{11}$})};
  \node (d) at (4,-4) {d({\color{blue}$\frac{3}{11}$})};
	\node (e) at (1,-6) {e({\color{blue}$\frac{3}{11}$})};
	\node (f) at (-2,-4) {f({\color{blue}$\frac{1}{11}$})}; 	
	\node (g) at (7,0) {g({\color{blue}$\frac{3}{11}$})};
	\node (h) at (10,-2) {h({\color{blue}$\frac{3}{11}$})};
	\node (i) at (10,-4) {i({\color{blue}$\frac{1}{11}$})};
	\node (j) at (10,-6) {j({\color{blue}$\frac{1}{11}$})};
	\node (k) at (7,-6) {k({\color{blue}$\frac{3}{11}$})}; 	
 \foreach \from / \to / \weight in {a/b/{\color{red}1/12},c/d/{\color{red}1/12},d/e/{\color{red}1/12},e/f/{\color{red}1/12},b/d/{\color{red}1/12},b/e/{\color{red}1/12},c/g/{\color{red}1/12},g/h/{\color{red}1/12},g/k/{\color{red}1/12},h/k/{\color{red}1/12},j/k/{\color{red}1/12},h/i/{\color{red}1/12}}
   \draw (\from) -- node[weight] {$\weight$} (\to);

	\end{tikzpicture}
	\caption{Example of assignment of \emph{normalized degrees} and \emph{initial edge weights}.}%
	\label{fig:local-density}
\end{figure}
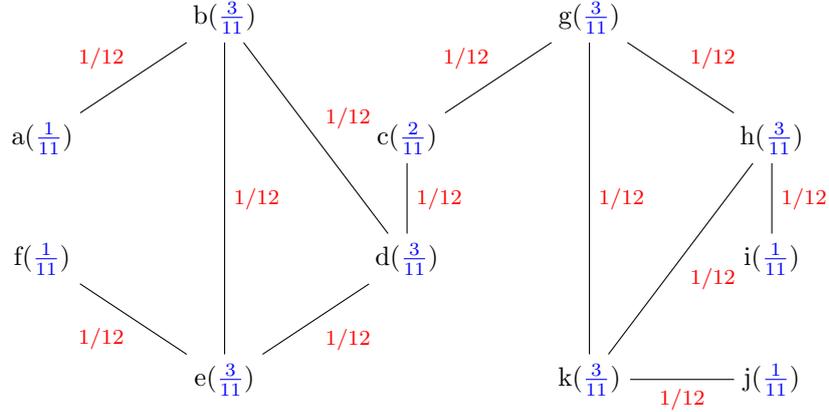

In Figure \ref{fig:local-density} we report an example of graph $G$ along with the distribution of
weights on nodes and edges.

\subsubsection{Step 2: Simulation of message propagations through random simple $\kappa$-paths}

In the second step we simulate multiple random walks on the graph $G$; this is consistent with Requirement 1.

To this purpose, our algorithm iterates the following sub-steps a number of times equal to a value
$\rho$, being $\rho$ a fixed value. We will later provide a practical rule for tuning $\rho$.
At each iteration, our algorithm performs the following operations:

\begin{enumerate}
		
\item A node $v_n \in V$ is selected according to one of the following two possible strategies:

				\begin{enumerate}
					\item[a.] uniformly at random, with a probability
					
					\begin{equation}
						P(v_n)= \frac{1}{|V|}
						\label{eq:pev}
					\end{equation}
					\item[b.] with a probability  proportional to its normalized degree $\delta(v_n)$, given by
					
					\begin{equation}
						\displaystyle{P(v_n)=\frac{\delta(v_n)}{\sum_{v_k \in V} \delta(v_k)}}
						\label{eq:ve}
					\end{equation}
					
				\end{enumerate}

\item  All the edges in $G$ are marked as not traversed.

\item The procedure {\em MessagePropagation} is invoked. It generates a simple random walk
    whose length is not greater than $\kappa$, satisfying Requirement 3.

\end{enumerate}

Let us describe the procedure {\em MessagePropagation}. This procedure carries out a loop as long
as {\em both} the following conditions hold true:

\begin{itemize}

\item The length of the path currently generated is no greater than $\kappa$.
			This is managed through a length counter $N$.

\item Assuming that the walk has reached the node $v_n$, there must exist at least an incident edge on $v_n$ which has not been already traversed.
			To do so, we attach a flag $T(e_m)$ to each edge $e_m \in E$, such that
			$$T(e_m) = \left\{
			\begin{aligned}
			&1 \quad  \mbox{if $e_m$  has   already  been  traversed}\\
			&0 \quad \mbox{otherwise}
			\end{aligned}\right.
			$$
			We observe that the following condition must be true
				\begin{equation}
					|I(v_n)| > \sum_{e_k \in I(v_n)}{T(e_k)}
					\label{eq:stop}
				\end{equation} being $I(v_n)$ the set of edges incident onto $v_n$.
				
\end{itemize}

The former condition complies with Requirement 4 (i.e., it allows us to consider only paths up to
length $\kappa$). The latter condition, instead, avoids that the message passes more than once
through an edge, thus satisfying Requirement 3.

If the conditions above are satisfied, the {\em MessagePropagation} procedure selects an edge $e_m$
by applying two strategies:
				
\begin{enumerate}
	\item[a.] uniformly at random, with a probability
								
		\begin{equation}
			P(e_m) = \frac{1}{|I(v_n)|- \sum_{e_k \in I(v_n)}{T(e_k)}}
			\label{eq:prob}
		\end{equation}
								
among all the edges $e_m \in \{I(v_n)\ | \ T(e_m) = 0\}$ incident on $v_n$ (i.e., excluding already traversed edges);
				
	\item[b.] with a probability proportional to the edge weight $\omega_l(e_m)$, given by

		\begin{equation}
			\displaystyle{P(e_m)=\frac{\omega_l(e_m)}{\sum_{e_m \in \hat{I}(v_n)} \omega_l(e_m)}}
			\label{eq:p-e-n}
		\end{equation}
					
	being $\hat{I}(v_n) = \{ e_k \in I(v_n) \ | \ T(e_k)=0 \}$ and
	\mbox{$\omega_l(e_m)=\omega_{l-1}(e_m)+\beta\cdot T(e_m)$} if $1\leq l \leq \kappa \rho$.
					
\end{enumerate}

Let $e_m$ be the selected edge and let $v_{n+1}$ be the node reached from $v_n$ by means of $e_m$.
The {\em MessagePropagation} procedure awards a bonus $\displaystyle{\beta}$ to $e_m$, sets $T(e_m)
= 1$ and increases the counter $N$ by 1. The message propagation activity continues from $v_{n+1}$.

At the end, each edge $e \in E$ is assigned a centrality index $L^\kappa(e)$ equal to its final weight $\omega_{\kappa \rho}(e)$.

The values of $\beta$ and $\rho$, in principle, can be fixed in an arbitrary fashion but we provide
a simple practical rule to tune them. Due to Theorem \ref{th:bounds}
reported in the Appendix, it emerges that in ERW-KPath it is convenient to set $\rho \simeq |E|$.
In particular, if we set $\rho = |E| - 1$ and $\beta = \frac{1}{|E|}$ we get a nice result: the
edge centrality indexes always range in $[\frac{1}{|E|}, 1]$ and, ideally, the centrality index of
a given edge will be equal to 1 if (and only if) it is {\em always} selected in any message
propagation simulation. In fact, each edge initially receives a default score equal to
$\frac{1}{|E|}$ and if that edge is selected in a subsequent trial, it will increase its score by a
factor $\beta = \frac{1}{|E|}$. Intuitively, if an edge is selected in all the trials, its final
score will be equal to $\frac{1}{|E|}+ \rho \cdot \frac{1}{|E|} = \frac{1}{|E|}+\frac{|E|-1}{|E|} =
1$.

The time complexity of this algorithm is $O(\kappa \rho)$. If we fix $\rho = |E| - 1$, we achieve a
good trade-off between accuracy and computational costs. In fact, in such a case, the worst case
time complexity of the ERW-KPath algorithm is $O(\kappa |E|)$ and, since in real social networks
$|E|$ is of the same order of magnitude of $|V|$, the time complexity of our approach is near
linear against the number of nodes. This makes our approach computationally feasible also for large
real-life social networks.

The version of the algorithm shown in Algorithms \ref{alg:ERW-Kpath} and \ref{alg:procedure} adopts
uniform probability distribution functions in order to choose nodes and edges purely at random and,
as said before, it is called ERW-KPath.

A weighted version of the same algorithm, called WERW-KPath, would differ only in line 5 (Algorithm
\ref{alg:ERW-Kpath}) and 2 (Algorithm \ref{alg:procedure}), adopting weighted functions specified
in Equations (\ref{eq:ve}) and (\ref{eq:p-e-n}). During our experimentation we always adopted the
WERW-Kpath algorithm, for the motivations explained in section \ref{sub:comparison}.

\begin{algorithm}
	\caption{ERW-Kpath(Graph $G=  \langle V,E \rangle$, int $\kappa$, int $\rho$, float $\beta$)}
	\label{alg:ERW-Kpath}
	\begin{algorithmic}[1]
		\STATE Assign each node $v_n \in V$ its normalized degree
		\STATE Assign each edge $e_m \in E$ the uniform probability function as weight
		\FOR{$i=1$ to $\rho$}			
			\STATE $N \leftarrow 0$ a counter to check the length of the $\kappa$-path
			\STATE $v_n \leftarrow$ a node chosen uniformly at random in $V$
			\STATE MessagePropagation($v_n$, $N$, $\kappa$, $\beta$)
		\ENDFOR
	\end{algorithmic}
\end{algorithm}

\begin{algorithm}
	\caption{MessagePropagation(Node $v_n$, int $N$, int $\kappa$, float $\beta$ )}
	\label{alg:procedure}
	\begin{algorithmic}[1]
				\WHILE{$N < \kappa$ and $\left[|I(v)| > \sum_{e \in I(v)}{T(e)}\right]$}
        \STATE $e_{m} \leftarrow$ $e_m \in \{I(v)\ | \ T(e_m) = 0\}$, chosen uniformly at random
				\STATE Let $v_{n+1}$ be the node reached by $v_n$ through $e_m$
        \STATE $\omega(e_{m}) \leftarrow \omega(e_{m}) + \beta$
				\STATE $T(e_{m}) \leftarrow  1$
				\STATE $v_n \leftarrow  v_{n+1}$
				\STATE $N \leftarrow N+1$
			\ENDWHILE
	\end{algorithmic}
\end{algorithm}

\subsection{Novelties introduced by our approach}
\label{sub:novelties}

In this section we discuss the main novelties introduced by our ERW-Kpath and WERW-Kpath algorithms.

First of all, we observe that our approach is {\em flexible} in the sense that it can be easily
modified to incorporate new models capable of describing the spread of a message in a network. For
instance, we can define multiple strategies to select the source node from which each message
propagation simulation starts. In particular, in this paper we considered two chances, namely:
\emph{(i)} the probability of selecting a node $s$ as the source is uniform across all the nodes in the
network (and this is at the basis of the ERW-Kpath algorithm) or \emph{(ii)} the probability of
selecting a node $s$ as the source is proportional to the degree of $s$ (and this is at the basis
of the WERW-Kpath). It would be easy to select a different probability distribution, if necessary.
In an analogous fashion, in the ERW-Kpath and WERW-Kpath algorithms we defined two strategies to
select the node receiving a message; of course, other, and more complex, strategies could be
implemented in order to replace those described in this paper.

In addition, observe that the ERW-Kpath and WERW-Kpath algorithms provide a {\em unicast
propagation model} in which any sender node is in charge of selecting {\em exactly one} receiving
node. We could easily modify our algorithms in such a way as to support a {\em multicast propagation
model} in which a node could issue a message to multiple receivers.

A further novelty is that we use multiple random walks to simulate the propagation of messages and
assume that the frequency of selecting an edge $e$ in these walks is a measure of its centrality.
An approach similar to our was presented in \cite{FoLaMa04} but it assumes that messages propagate
along shortest paths. In detail, given a pair of nodes $i$ and $j$, the approach of \cite{FoLaMa04}
introduces a parameter, called {\em network efficiency} $\varepsilon_{ij}$ as the inverse of the
length of the shortest path(s) connecting $i$ and $j$. After that, it provides a new parameter, called
{\em information centrality}; the information centrality $IC_e$ of an edge $e$ is defined as the
relative drop in the network efficiency generated by the removal of $e$ from the network. Our
approach provides some novelties in comparison with that of \cite{FoLaMa04}: in fact, in our
approach a network is viewed as a {\em decentralized system} in which there is no user having a
complete knowledge of the network topology. Due to this incomplete knowledge, users are
not able to identify shortest path and, therefore, they use a probabilistic model to spread
messages. This yields also relevant computational consequences: the identification of all the pairs of
shortest paths in a network is computationally expensive and it could be unfeasible on networks containing millions of nodes. By contrast, our approach scales almost linearly with the number of edges and, therefore, it can easily run
also over large networks.

Finally, despite our approach relies on the concept of message propagation which requires an
orientation on edges, it can work also on {\em undirected networks}. In fact, the ERW-Kpath (resp.,
WERW-Kpath) algorithm selects at the beginning a source node $s$ that decides the node $v$ to which
a message has to be forwarded. Therefore, at {\em run-time}, the ERW-Kpath (resp., WERW-Kpath)
algorithm {\em induces} an orientation on the edge linking $s$ and $v$ which coincides with the
direction of the message sent by $s$; such a process does not require to operate on directed
networks, even if it could intrinsically work well with such a type of networks.

\subsection{Comparison of the ERW-Kpath and WERW-Kpath algorithms}\label{sub:comparison}

In this section we provide a comparison between ERW-Kpath and WERW-Kpath. First of all, we
would like to observe that, according to Theorem \ref{th:bounds}, both the two algorithms are
capable of correctly approximating the $\kappa$-path centrality values provided in Definition
\ref{def:k-path-edge-centrality}.

Despite the two algorithms are formally correct, however, we observe that the WERW-Kpath algorithm
should be preferred to ERW-Kpath. In fact, in the ERW-Kpath algorithm, we assume that each
node can select, {\em at random}, any edge (among those that have not yet been selected) to
propagate a message. Such an assumption could be, however, too strong in real-life social networks.
To better clarify this concept, consider online social networks like Facebook or Twitter. In both
of these networks a single user may have a large number of contacts with whom she/he can exchange
information (e.g., a \emph{wall post} on Facebook or a \emph{tweet} on Twitter). However,
sociological studies reveal that there is an upper limit to the number of people with whom a user
could maintain stable social relationships and this number is known as {\em Dunbar number}
\cite{dunbar1998grooming}. For instance, in Facebook, the average number of friends of a user is
120. On the other hand, it has been reported that male users actively communicate with only 10 of
them, whereas female users with 16\footnote{{\tt
http://www.economist.com/node/13176775?story\_id=13176775}}. This implies that there are
preferential edges along which information flows in social networks.

The ERW-Kpath algorithm is simple and easy to implement but it could fail to identify preferential edges
along which messages propagate. By contrast, in the WERW-Kpath algorithm, the probability of
selecting an edge is proportional to the weight already acquired by that edge. This weight,
therefore, has to be intended as the frequency with which two nodes exchanged messages in the past.

Such a property has also a relevant implication and makes feasible some applications which could
not be implemented by the ERW-Kpath algorithm. In fact, our approach, to some extent can be
exploited to recommend/predict links in a social network. The problem of recommending/predicting
links plays a key role in Computer Science and Sociology and it is often known in the literature as
the {\em link prediction problem} \cite{liben2007link}. In the link prediction problem, the
network topology is analyzed to find pairs of non-connected nodes which could get a profit by
creating a social link. Various measures can be exploited to assess whether a link should be
recommended between a pair of nodes $u$ and $v$; for instance, the simplest measure is to compute
the Jaccard coefficient $J(u,v)$ on the neighbors of $u$ and $v$. The larger the number of
neighboring nodes shared by $u$ and $v$, the larger $J(u,v)$; in such a case it is convenient to
add an edge in the network linking $u$ and $v$. Further (and more complex measures) take the {\em
whole network topology} into account to recommend links. For instance, the {\em Katz coefficient}
\cite{liben2007link} considers the whole ensemble of paths running between $u$ and $v$ to decide
whether a link between them should be recommended.

The WERW-Kpath algorithm can be exploited to address the link prediction problem. In detail, by means
of WERW-Kpath, we can handle not only {\em topological information} but we can also
quantify the strength of the relationship joining two nodes. So, we know that two nodes $u$ and $v$
are connected and, in addition, we know also how frequently they exchange information. This allows
us to extend the measure introduced above: for instance, if we would like to use the Jaccard
coefficient, we can consider only those edges (called {\em strong edges}) coming out from $u$ (resp.,
$v$) such that the weight of these edge is greater than a given threshold. This is equivalent to
filter out all the edges which are rarely employed to spread information. As a consequence, the
Jaccard coefficient could be computed only on strong edges.

Due to these reasons, in the following experiments we focused only on the WERW-Kpath algorithm.

\section{Experimentation} \label{sec:experimentation}
Our experimentation has been conducted on different online social networks whose datasets are available.
Adopted datasets have been summarized in Table \ref{tab:datasets}.

Dataset 1 depicts the voting system of Wikipedia for the elections of January 2008. 
Datasets 2 and 3 represent the Arxiv\footnote{Arxiv (http://arxiv.org/) is an online archive for scientific preprints in the fields of Mathematics, Physics and Computer Science, amongst others.} archives of papers in the field of, respectively, High Energy Physics (Phenomenology) and Condensed Matter Physics, as of April 2003. 
Dataset 4 represents a network of scientific citations among papers belonging to the Arxiv High Energy Physics (Theory) field. 
Dataset 5 describes a small sample of the Facebook network, representing its friendship graph. 
Finally, Dataset 6 depicts a fragment of the YouTube social graph as of 2007.

\begin{table}
\centering
\footnotesize
\begin{tabular}{c c c c c c c c}
	\hline
	\#	&	Network 			&	No. Nodes	\	&	\ No. Edges	\	&	\ Directed	\			&	Type 					& \ Ref \ \\	
	\hline \hline
	1	&	Wiki-Vote			& 7,115			&	103,689		&	Yes			&	Elections			&	\cite{snap}			\\
	2	&	CA-HepPh			&	12,008		&	237,010		&	No 			& Co-authors		&	\cite{snap}			\\
	3	&	CA-CondMat		&	23,133		&	186,932		& No			&	Co-authors		&	\cite{snap}			\\	
	4	&	Cit-HepTh			&	27,770		&	352,807		&	Yes			& Citations			& \cite{snap}			\\
	5	&	Facebook			&	63,731		&	1,545,684	&	Yes			& Online SN			& \cite{viswanath2009evolution}\\
	6	& Youtube				&	1,138,499	&	4,945,382	&	No			& Online SN			& \cite{viswanath2009evolution}\\
	\hline
\end{tabular}
\caption{Datasets adopted in our experimentation.}
\label{tab:datasets}
\end{table}

\subsection{Robustness} \label{sec:reproducibility}
A quality required for a good random-walk based algorithm is the \emph{robustness of results}.
In fact, it is important that obtained results are consistent among different iterations of the algorithm, if initial conditions are the same.
In order to verify that our WERW-Kpath produces reliable results, we performed a \emph{quantitative} and a \emph{qualitative} analysis as follows.

In the quantitative analysis we are interested in checking whether the algorithm produces the same
results in different runs. In the qualitative analysis, instead, we studied whether different
values of $\kappa$ deeply impact on the ranking of edges.

\subsubsection{Quantitative analysis of results} \label{sec:quantitative}
Our first experimentation is in order to verify that, over different iterations with the same configuration, results are consistent.
It is possible to highlight this aspect, running several times the WERW-Kpath algorithm on the same dataset, with the same configuration.

Regarding $\rho$, in the experimentation we adopt $\rho = |E| - 1$, which is consistent with
Theorem \ref{th:bounds}. According to the previous choice, the bonus awarded is fixed to $\beta =
\frac{1}{|E|}$. As for the maximum length of the $\kappa$-paths, we chose a value of $\kappa = 20$.

Our quantitative analysis highlights that the distributions of values are almost completely
overlapping, over different runs on each dataset among those considered in Table
\ref{tab:datasets}.

\begin{figure}[!ht]%
	\centering
	\includegraphics[width=.6\columnwidth]{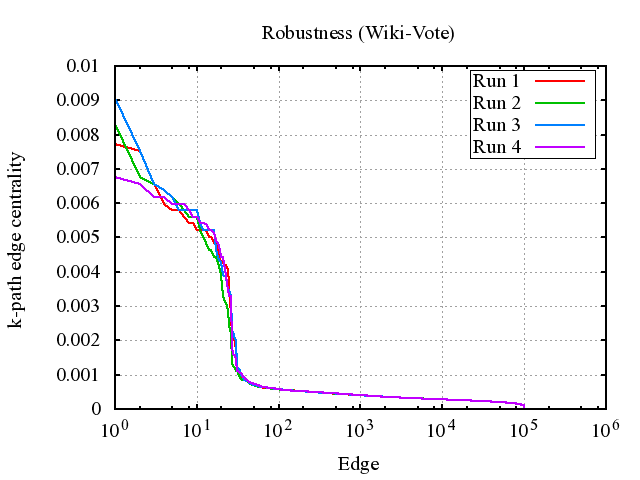}%
	\caption{Robustness test on ``Wiki-Vote''.}%
	\label{fig:reproducibility}%
\end{figure}

In Figure \ref{fig:reproducibility} we graphically report the distribution of edge centrality
values for the ``Wiki-Vote'' dataset. Results are from four different runs of the algorithm on the
same dataset with the same configuration. Data are plotted using a semi-logarithmic scale in
order to highlight the ``high'' part of the distribution, where edges with high $\kappa$-path edge
centrality lie.

Similar results are confirmed performing the same test over each considered dataset but they are
not reported due to space limitations. The robustness property is necessary but not sufficient to
ensure the correctness of our algorithm.

In fact, the quantitative evaluation we performed ensures that centrality values produced by WERW-Kpath are consistent over different runs of the algorithm, but does not ensure that, for example, a same edge $e \in E$ after the \emph{Run 1} has a centrality value which is the same (or, at least, very similar) that after \emph{Run 2}.
In other words, those values of centrality that overlap in different distributions may be not referred to the same edges.

To the purpose of investigating this aspect we analyze results from a qualitative perspective, as follows.

\subsubsection{Qualitative analysis of results} \label{sec:qualitative}

Our random-walk-based approach ensures minimum fluctuations of centrality values assigned to each edge along different runs, if the configuration of each run is the same.

To verify this aspect, we calculate the similarity of the distributions obtained by running
WERW-Kpath four times on each dataset, using the same configuration, comparing results by adopting different measures.
For this experiment, we considered different settings for the length of the exploited $\kappa$-paths, i.e., $\kappa = 5, 10, 20$, in order to investigate also its impact.

The first measure considered is a variant of the \emph{Jaccard coefficient}, classically defined as

\begin{equation}
	J(X,Y) = \frac{|X \cap Y|}{|X \cup Y|}
	\label{eq:jaccard}
\end{equation}
where $X$ and $Y$ represent, in our case, a pair of compared distributions of $\kappa$-path edge centrality values.

In order to define the Jaccard coefficient in our context we need to take into account the
following considerations. Let us consider two runs of our algorithms, say $X$ and $Y$ and let us
first consider an edge $e$; let us denote with $\omega_X(e)$ (resp., $\omega_Y(e)$) the centrality index
of $e$ in the run $X$ (resp., $Y$); intuitively, the performance of our algorithm is ``good'' if
$\omega_X(e)$  is close to $\omega_Y(e)$; however, a direct comparison of the two values could make
no sense because, for instance, the edge $e$ could have the highest weight in both the two runs but
$\omega_X(e)$ may significantly differ from $\omega_Y(e)$. Therefore, we need to consider the {\em
normalized values} $\frac{\omega_X(e)}{\max_{e \in X} \omega(e)}$ and $\frac{\omega_Y(e)}{\max_{e
\in Y}\omega(e)}$ and we assume that the algorithm yields good results if these values are
``close''. To make this definition more rigorous we can define $\Lambda(e) = \left|
\frac{\omega_X(e)}{\max_{e \in X}\omega(e)} - \frac{\omega_Y(e)}{\max_{e \in Y}\omega(e)} \right|$
and we say that the algorithm produces good results if $\Lambda(e)$ is smaller than a threshold
$\varepsilon$.

Now, in order to fix the value of $\varepsilon$, let us consider the values achieved by
$\Lambda(e)$ for each $e \in E$. We can provide an upper bound $\overline{\Lambda}$ on $\Lambda(e)$
by considering two extremal cases: \emph{(i)} $\omega_X(e) = \max_{e \in X}\omega(e)$ and
$\omega_Y(e) = \min_{e \in Y}\omega(e)$ or, vice versa, \emph{(ii)} $\omega_X(e) = \min_{e \in
X}\omega(e)$ and $\omega_Y(e) = \max_{e \in Y}\omega(e)$. For the sake of simplicity, assume that
case \emph{(i)} occurs; of course, the following considerations hold true also in case \emph{(ii)}.
In such a case we obtain $\overline{\Lambda} = \left| 1 - \frac{\min_{e \in Y} \omega(e)}{\max_{e
\in Y}\omega(e)} \right|$. As discussed in the following (see Figures \ref{fig:k-paths} and
\ref{fig:k-correlation}), edge centralities are distributed according to a power law and,
therefore, the value of $\min_{e \in Y} \omega(e)$ is some orders of magnitude smaller than
$\max_{e \in Y}\omega(e)$. Therefore, the ratio of $\min_{e \in Y}\omega(e)$ to $\max_{e \in
Y}\omega(e)$ tends to 0 and $\overline{\Lambda}$ tends 1.

According to these considerations, we computed how many times the following condition holds true
$\Lambda(e) \leq \tau \overline{\Lambda}$, being $ 0 < \tau \leq 1 $ a {\em tolerance threshold}.
Since $\overline{\Lambda} \simeq 1$, this amounts to counting how many times $\Lambda(e) \leq
\tau$. Therefore, we can define the {\em modified Jaccard coefficient} as follows

\begin{equation}
	J^\tau(X,Y) =  \frac{|\{e: | \frac{\omega_X(e)}{\max_{e \in X}\omega(e)} - \frac{\omega_Y(e)}{\max_{e \in Y}\omega(e)}  | \leq \tau \}|}{|X \cup Y|}
	\label{eq:tol-coeff}
\end{equation}

In our tests we considered the following values of tolerance $\tau=0.01,0.05,0.10$ to identify 1\%, 5\% and 10\% of maximum accepted variation of the edge centrality value assigned to a given edge along different runs with same configurations.

A mean degree of similarity $avg\left[J_{\binom{n}{k}}^\tau \right]$ is taken to average the
\mbox{$\binom{4}{2} = 6$} possible combinations of pairs  of distributions
obtained by analyzing the four runs over the datasets discussed above.

The second measure we consider is the \emph{Pearson correlation}.
It is adopted to evaluate the correlation of the two obtained distributions.
It is defined as

\begin{equation}
	\rho_{X,Y} = \frac{cov(X,Y)}{\sqrt{var(X)\cdot var(Y)}}
\label{eq:pearson}
\end{equation}

whose results are normalized in the interval $[-1,+1]$, with the following interpretations:
\begin{itemize}
	\item $\rho_{X,Y} > 0$: distributions are directly correlated, in particular:
		\begin{itemize}
			\item $\rho_{X,Y} > 0.7$: strongly correlated;
			\item $0.3 < \rho_{X,Y} < 0.7$: moderately correlated;
			\item $0 < \rho_{X,Y} < 0.3$: weakly correlated;
		\end{itemize}
	\item $\rho_{X,Y} = 0$: not correlated;
	\item $\rho_{X,Y} < 0$: inversely correlated.
\end{itemize}

Clearly, the higher $\rho_{X,Y}$, the better the WERW-KPath algorithm works.
Observe that the $\rho_{X,Y}$ coefficient tells us whether the two distributions $X$ and $Y$ are deterministically related or not.
Therefore, it could happen that the WERW-KPath algorithm, in two different runs generates two edge centrality distributions $X$ and $Y$ such that $Y = aX$, being $a$ a real coefficient. In such a case, the $\rho_{X,Y}$ coefficient would be 1 but we could not conclude that the algorithm works properly. In fact, the coefficient $a$ could be very low (or in the opposite case very large) and, therefore, the two distributions would significantly differ even if they would preserve the same edge rankings.

To this purpose, we consider a third measure in order to compute the distance between the two distributions $X$ and $Y$.
To do so, we adopt the Euclidean distance $L_2(X,Y)$ defined as

\begin{equation}
L_2(X,Y)=\sqrt{\sum_{i=1}^{n}{(X_i - Y_i)}^2}
\label{eq:l2}
\end{equation}

As it emerges from the distributions shown in Figure \ref{fig:reproducibility}, almost all the terms in Equation (\ref{eq:l2})
annul each other, and therefore, the final value of $L_2(X,Y)$ is dominated by the difference of
the $\kappa$-path centrality values associated with the few top-ranked edges. To obtain the average
distance between two points in distribution $X$ and $Y$ in a given dataset, we should simply divide
$L_2(X,Y)$ by the number of edges in that dataset.

Intrinsic characteristics of analyzed datasets do not influence the robustness of
results. In fact, even if considering datasets representing different social networks (e.g.,
collaboration networks, citation networks and online communities), WERW-Kpath produces highly
overlapping results over different runs.

Already adopting a low tolerance, such as $\tau=0.01$ or $\tau=0.05$, values of $\kappa$-path edge centrality are highly overlapping.
Results improve according to the length of the $\kappa$-path adopted.
By increasing tolerance and/or length of $\kappa$-paths, the full overlap became obvious.
The same considerations hold true with respect to the Pearson correlation coefficient which identifies strong correlations among all the different distributions.

Finally, as for the Euclidean distance, we observe that returned values are always small and, in
every case the distance is no larger than $[10^{-2}, 10^{-3}]$ and the average distance is around
$[10^{-7}, 10^{-10}]$.

\begin{table}
\footnotesize
\centering
	\begin{tabular}{c c c c c c c c}
	\cline{3-5}
	\multicolumn{2}{c}{} & \multicolumn{3}{|c|}{$\displaystyle{J_{\binom{n}{k}}^{\tau}}$} & \multicolumn{2}{c}{}  \\
	\hline
	Dataset &	$\kappa$ 	& 	$\tau=0.01\ $ &	$\tau=0.05\ $ &	$\tau=0.10\ $  & $\rho_{X,Y}$ & $L_2(X,Y)$ & avg($L_2(X,Y)$)\\
	\hline
	\hline

	\multirow{3}{*}{\small Wiki-Vote}	& $\kappa$ = 5 & 43.52\% &98.49\%	&	99.91\%	& 0.67	&	1.61$\cdot 10^{-2}$	& 1.55$\cdot 10^{-7}$\\
																		&	$\kappa$ = 10& 61.13\% &98.86\%	&	99.98\%	&	0.69	&	2.37$\cdot 10^{-2}$ & 2.28$\cdot 10^{-7}$\\
																		&	$\kappa$ = 20& 70.68\% &99.96\%	&	99.98\%	& 0.70	&	3.48$\cdot 10^{-2}$ & 3.35$\cdot 10^{-7}$\\
	\hline

	\multirow{3}{*}{\small CA-HepPh}	& $\kappa$ = 5 & 52.63\% &96.11\%	&	99.53\% &	0.92	&	1.18$\cdot 10^{-2}$ & 4.97$\cdot 10^{-8}$\\
																		&	$\kappa$ = 10& 70.45\% &99.02\% &	99.88\% &	0.95	&	1.23$\cdot 10^{-2}$ & 5.18$\cdot 10^{-8}$\\
																		&	$\kappa$ = 20& 75.65\% &99.51\%	&	99.87\%	& 0.96	&	2.90$\cdot 10^{-2}$ & 1.22$\cdot 10^{-7}$\\	
	\hline
	
	\multirow{3}{*}{\small CA-CondMat}& $\kappa$ = 5 & 22.23\% &80.51\%	&	96.98\%	&	0.73	&	1.39$\cdot 10^{-2}$ & 7.43$\cdot 10^{-8}$\\
																		&	$\kappa$ = 10& 35.16\% &93.72\%	&	99.40\% &	0.79	&	2.18$\cdot 10^{-2}$ & 1.16$\cdot 10^{-7}$\\
																		&	$\kappa$ = 20& 35.63\% &95.80\%	&	99.44\%	&	0.83	&	3.40$\cdot 10^{-2}$ & 1.81$\cdot 10^{-7}$\\
	\hline

	\multirow{3}{*}{\small Cit-HepTh}	& $\kappa$ = 5 & 47.62\% &97.76\%	&	99.78\%	&	0.78	&	0.92$\cdot 10^{-2}$ & 2.60$\cdot 10^{-8}$\\  	
																		&	$\kappa$ = 10& 60.61\% &99.45\%	&	99.93\%	&	0.83	&	1.36$\cdot 10^{-2}$ & 3.85$\cdot 10^{-8}$\\
																		&	$\kappa$ = 20& 63.68\% &99.62\%	&	99.93\%	&	0.85	&	2.04$\cdot 10^{-2}$ & 5.78$\cdot 10^{-8}$\\

	\hline
	\multirow{3}{*}{\small Facebook}	&	$\kappa$ = 5	& 56.98\% &	97.34\%	&	99.36\%	&	0.79	&	1.01$\cdot 10^{-2}$ & 5.11$\cdot 10^{-9}$\\
																		& $\kappa$ = 10 & 56.85\% &	98.49\%	&	99.76\%	&	0.84	&	1.87$\cdot 10^{-2}$ & 1.20$\cdot 10^{-8}$\\ 	
																		&	$\kappa$ = 20 & 68.58\% &	99.39\%	&	99.90\%	&	0.84	&	2.67$\cdot 10^{-2}$ & 1.72$\cdot 10^{-8}$\\
	
	\hline																		
	\multirow{3}{*}{\small Youtube}		&	$\kappa$ = 5	& 11.74\% &	44.28\%	&	72.41\%	&	0.49	&	1.31$\cdot 10^{-3}$ & 2.64$\cdot 10^{-10}$\\
																		& $\kappa$ = 10 & 13.18\% &	59.40\%	&	84.91\%	&	0.75	&	1.87$\cdot 10^{-3}$ & 3.78$\cdot 10^{-10}$\\ 	
																		&	$\kappa$ = 20 & 27.92\% &	82.29\%	&	96.17\%	&	0.89	&	2.83$\cdot 10^{-3}$ & 5.72$\cdot 10^{-10}$\\

	\hline	\hline																								
	\end{tabular} \caption{Analysis by using similarity coefficient
$\displaystyle{J_{\binom{n}{k}}^{\tau}}$, correlation $\rho_{X,Y}$ and Euclidean distance
$L_2(X,Y)$.} \label{tab:jaccard}
\end{table}

\subsection{Performance} \label{sec:performance}
All the experiments have been carried out by using a standard Personal Computer equipped with a Intel i5 Processor with 4 GB of RAM.
The implementation of the WERW-Kpath algorithm adopted in the following experiments, developed by using Java 1.6, has been released\footnote{http://www.emilio.ferrara.name/werw-kpath/} and its adoption is strongly encouraged.

\begin{figure}[!ht]
	\centering
	\includegraphics[width=.6\columnwidth]{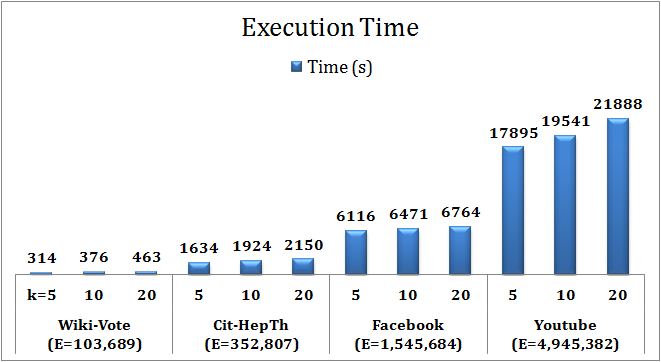}%
	\caption{Execution time with respect to network size.}%
	\label{fig:k-execution-time}%
\end{figure}

As shown in Figure \ref{fig:k-execution-time}, the execution of WERW-Kpath scales very well (i.e., almost linearly) according with the setup of the length of the $\kappa$-paths and with respect to the number of edges in the given network.

This means that this approach is feasible also for the analysis of large networks, making it
possible to compute an efficient centrality measure for edges in all those cases in which it would
be very difficult or even unfeasible, for the computational cost, to calculate the exact edge-betweenness \cite{girvan2002community}.

The importance of this aspect is evident if we consider that there exist several Social Network Analysis tools, that implement different algorithms to compute centrality indices on network nodes/edges.
Our measure could be integrated in such tools (e.g., NodeXL\footnote{http://nodexl.codeplex.com/}, Pajek\footnote{http://pajek.imfm.si/doku.php?id=pajek}, NWB\footnote{http://nwb.cns.iu.edu/}, and so on), in order to allow social network analysts, to manage (possibly, even larger) social networks in order to study the centrality of edges.

\subsection{Analysis of Edge Centrality Distributions} \label{sec:results}

In this section we study the distribution of edge centrality values computed by the WERW-Kpath algorithm.
In detail, we present the results of two experiments.

\begin{figure}[!ht] \centering
		\includegraphics[width=.45\columnwidth]{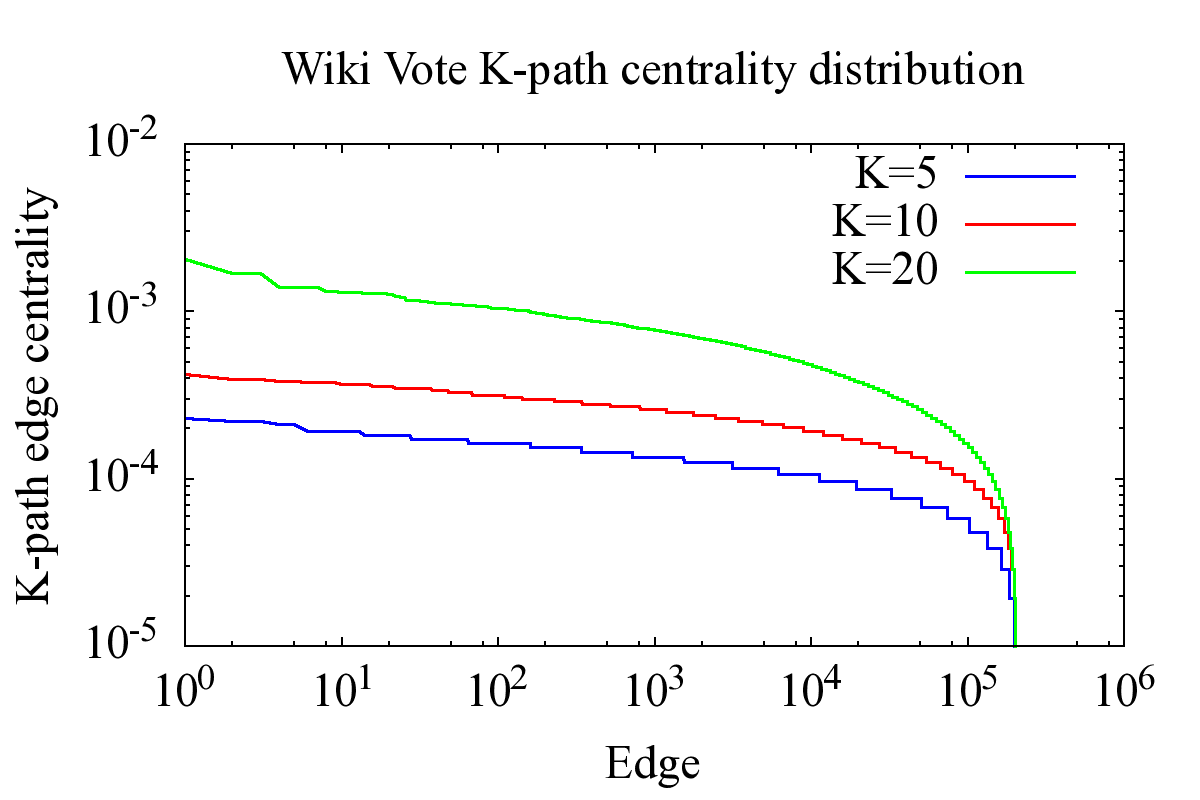}%
		\includegraphics[width=.45\columnwidth]{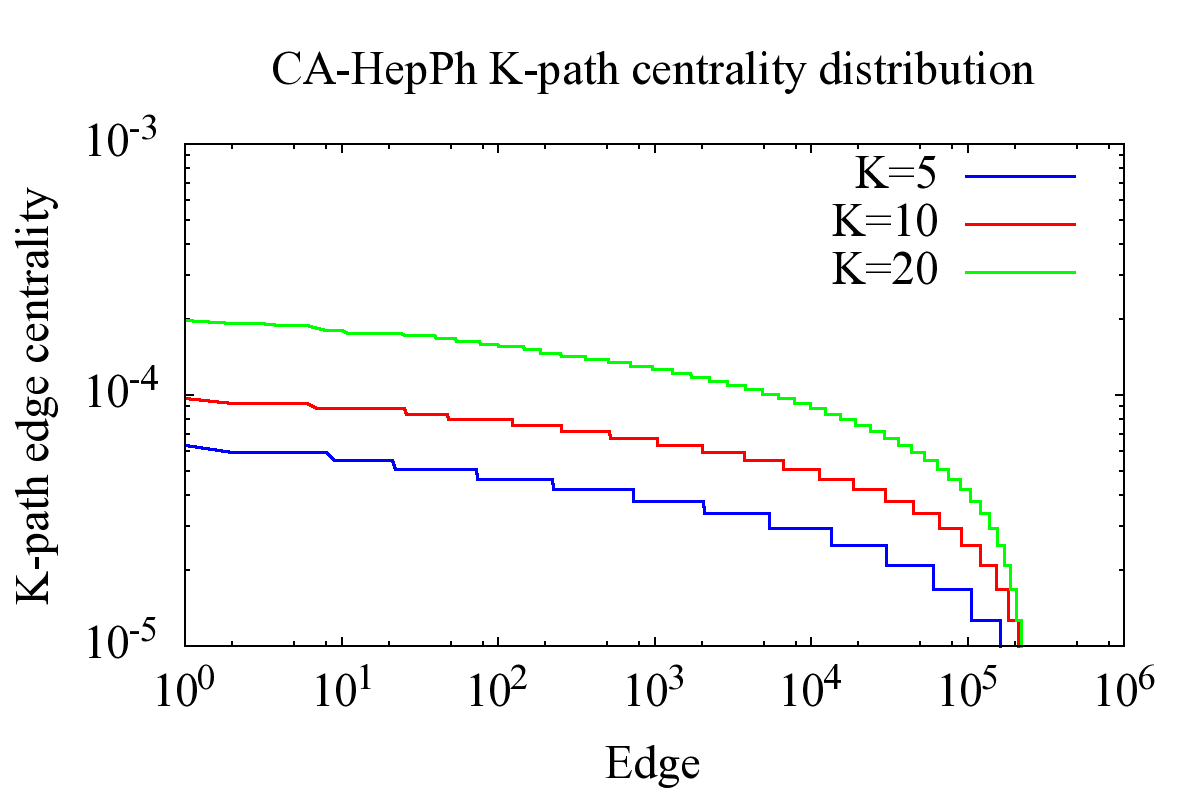}
		\includegraphics[width=.45\columnwidth]{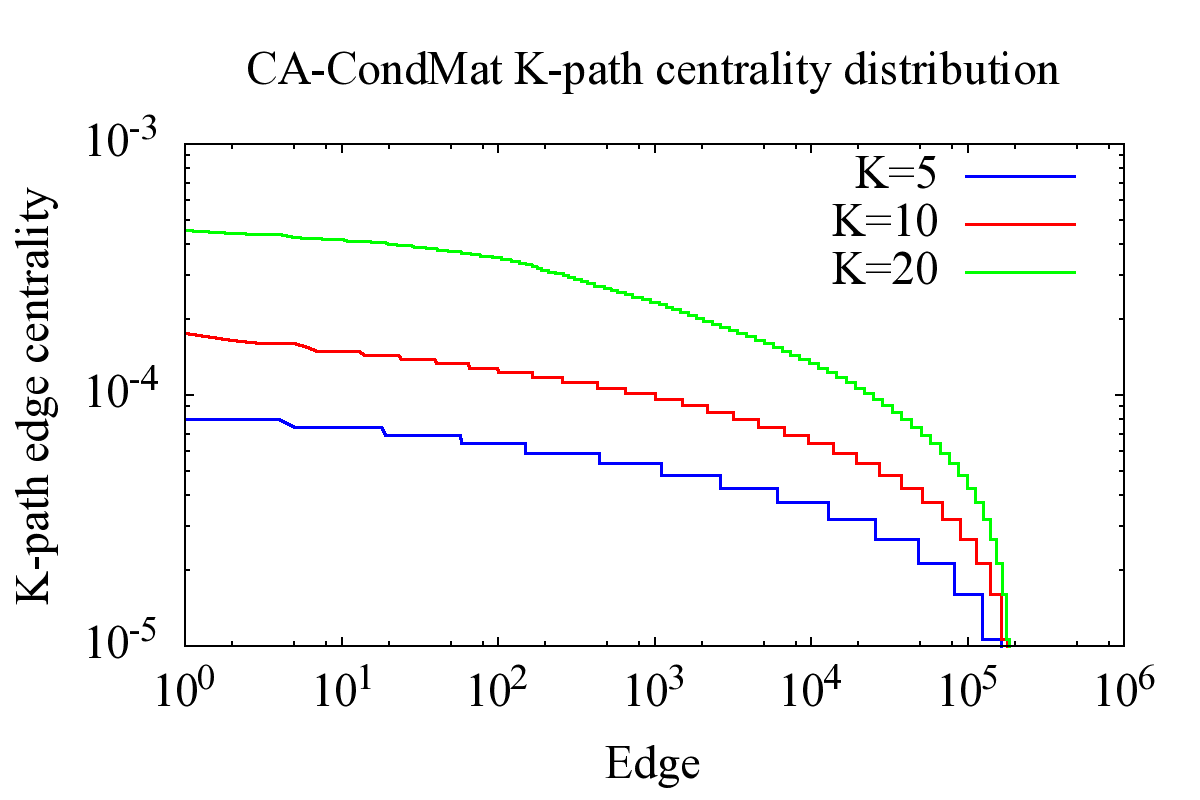}%
		\includegraphics[width=.45\columnwidth]{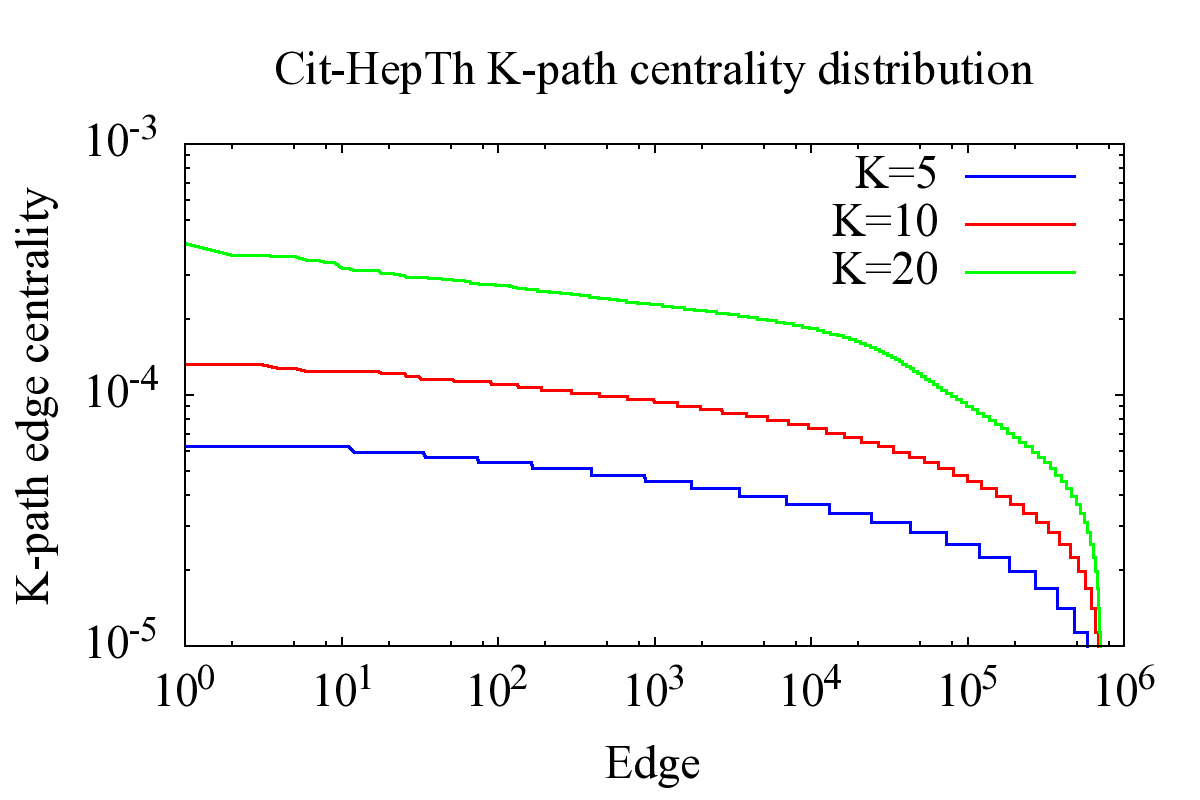}
		\includegraphics[width=.45\columnwidth]{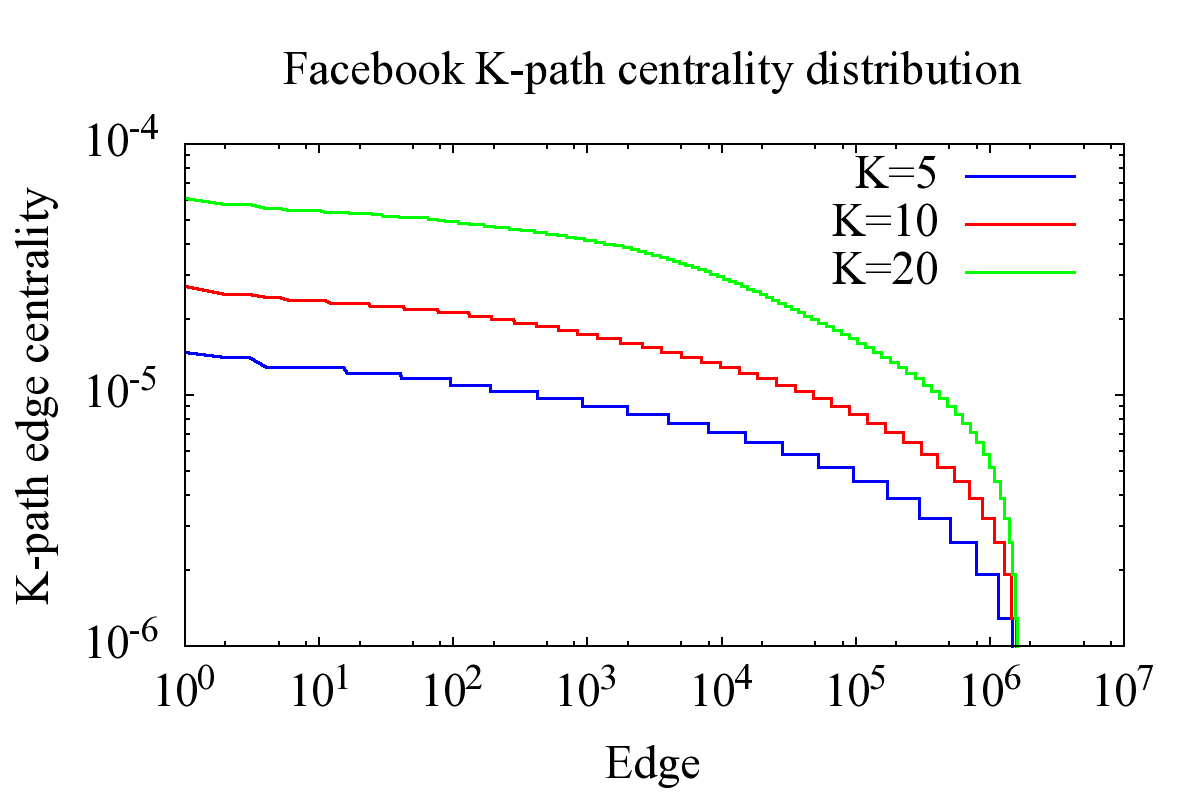}
		\includegraphics[width=.45\columnwidth]{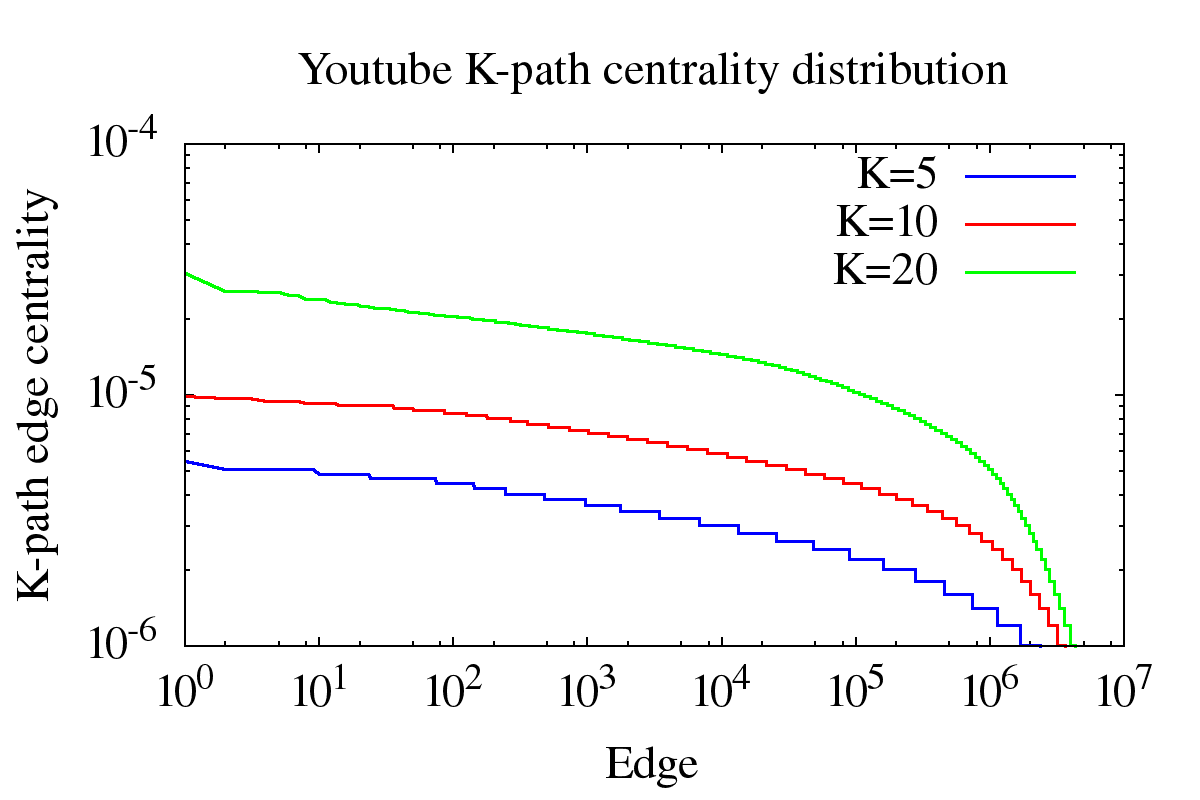}
	\caption{$\kappa$-paths centrality values distribution on different networks.}%
	\label{fig:k-paths}%
\end{figure}

In the first experiment we ran our algorithm four times. In addition, we varied the value of
$\kappa = 5, 10, 20$. We averaged the $\kappa$-path centrality values at each iteration and we
plotted the edge centrality distribution; on the horizontal axis we reported the identifier of each
edge. The results are reported in Figure \ref{fig:k-paths} by exploiting a logarithmic scale. The
figure has the following interpretation: on the x-axis it represents each edge of the given
network, on the y-axis its corresponding value of $\kappa$-path edge centrality.

The usage of a logarithmic scale highlights a power law distribution for the centrality values. In
fact, when the behavior in a log-log scale resembles a straight line, the distribution could be
well approximated by using a power law function $f(x) \propto x^{-\alpha}$. As a result, for the all
considered datasets, there are few edges with high centrality values whereas a large fraction of
edges presents low (or very low) centrality values. Such a result can be explained by recalling
that, at the beginning, our algorithm considers all the edges on an equal foot and provides them with
an initial score which is the same for all the edges. However, during the algorithm execution, it
happens that few edges (which are actually the \emph{most central} edges in a social network) are
frequently selected and, therefore, their centrality index is frequently updated. By contrast, many
edges are seldom selected and, therefore, their centrality index is rarely increased. This process
yields a power law distribution in edge centrality values.

\begin{figure}[!ht] \centering
		\includegraphics[width=.45\columnwidth]{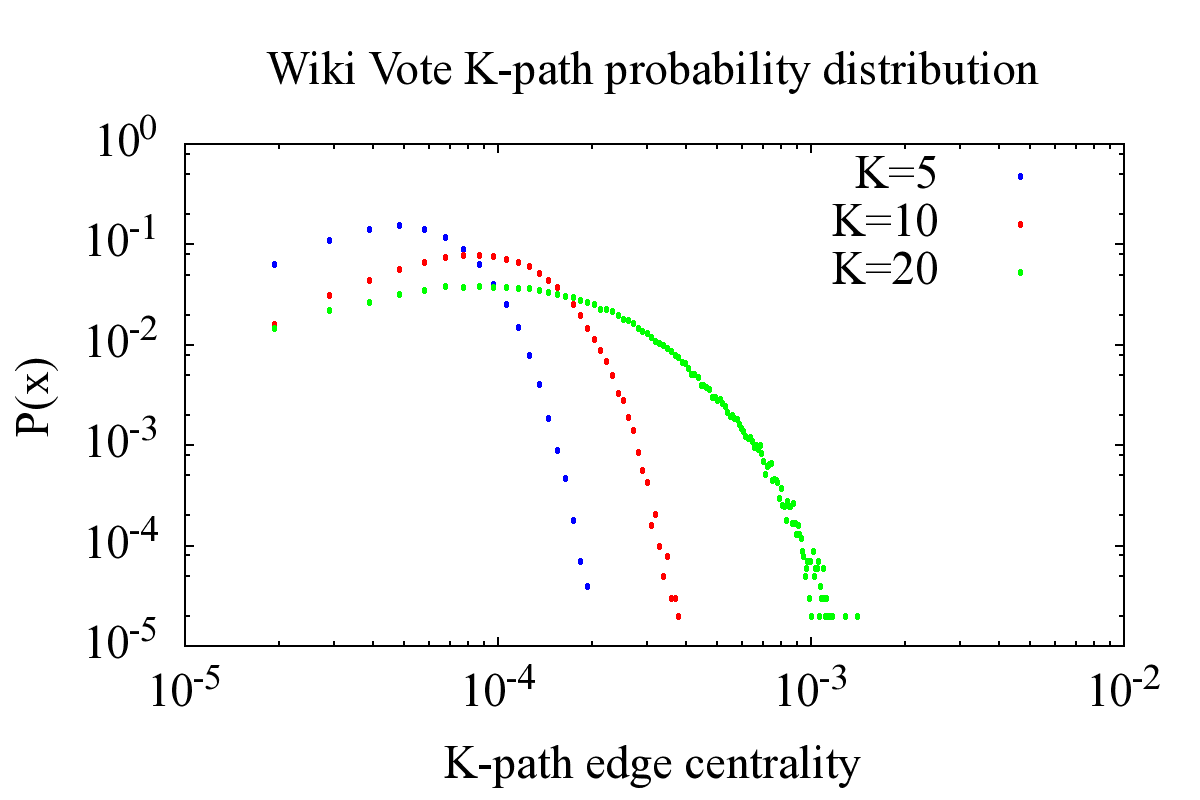}%
		\includegraphics[width=.45\columnwidth]{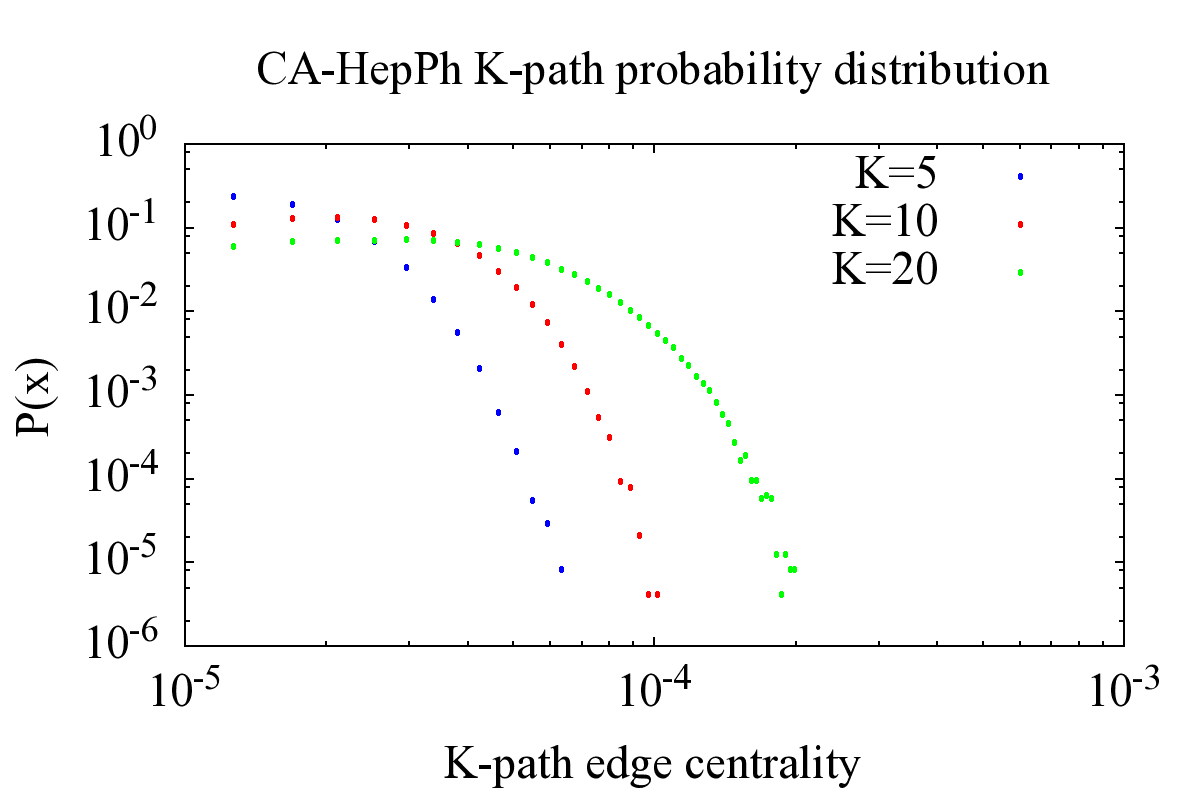}
		\includegraphics[width=.45\columnwidth]{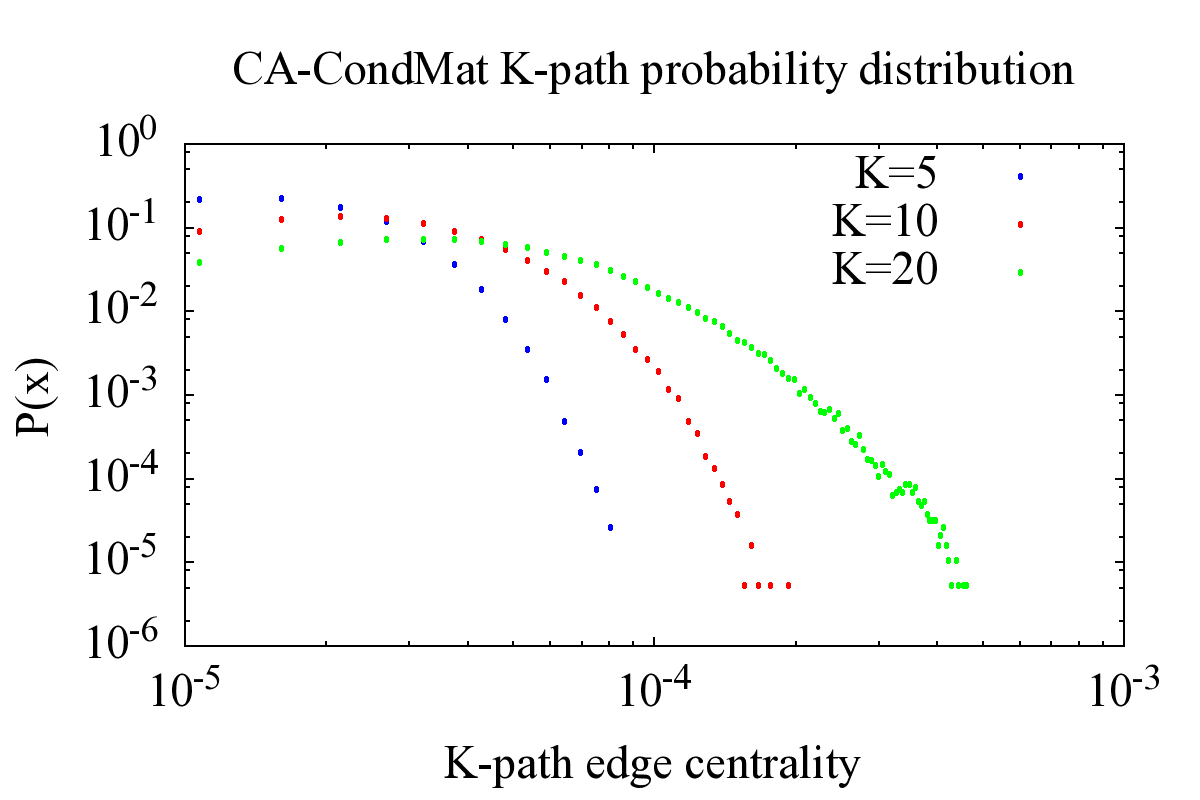}%
		\includegraphics[width=.45\columnwidth]{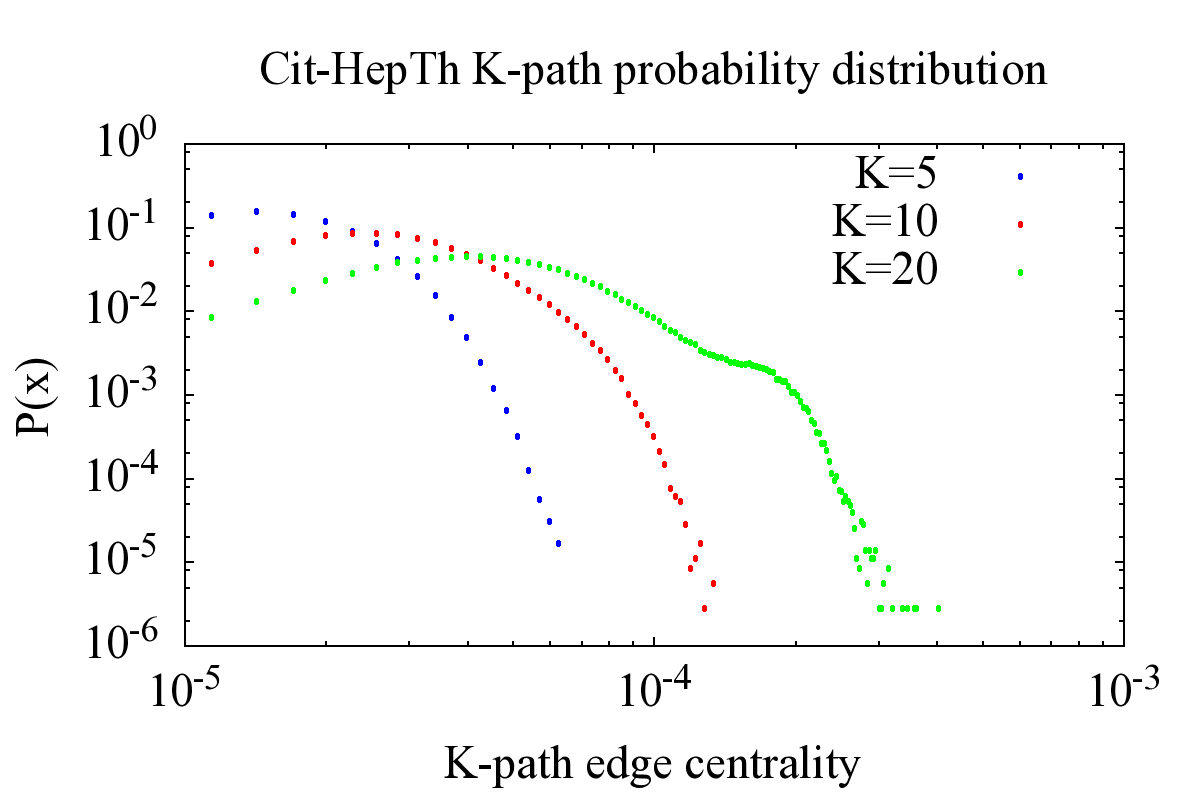}
		\includegraphics[width=.45\columnwidth]{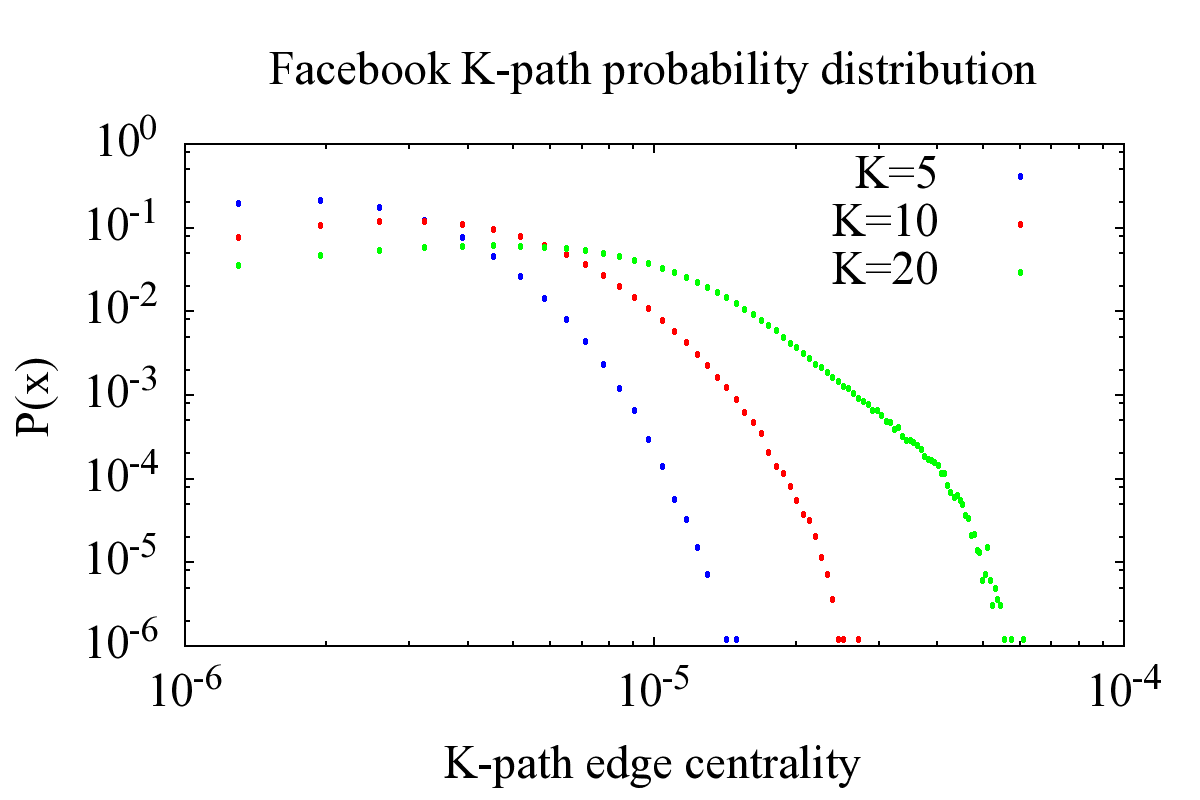}
		\includegraphics[width=.45\columnwidth]{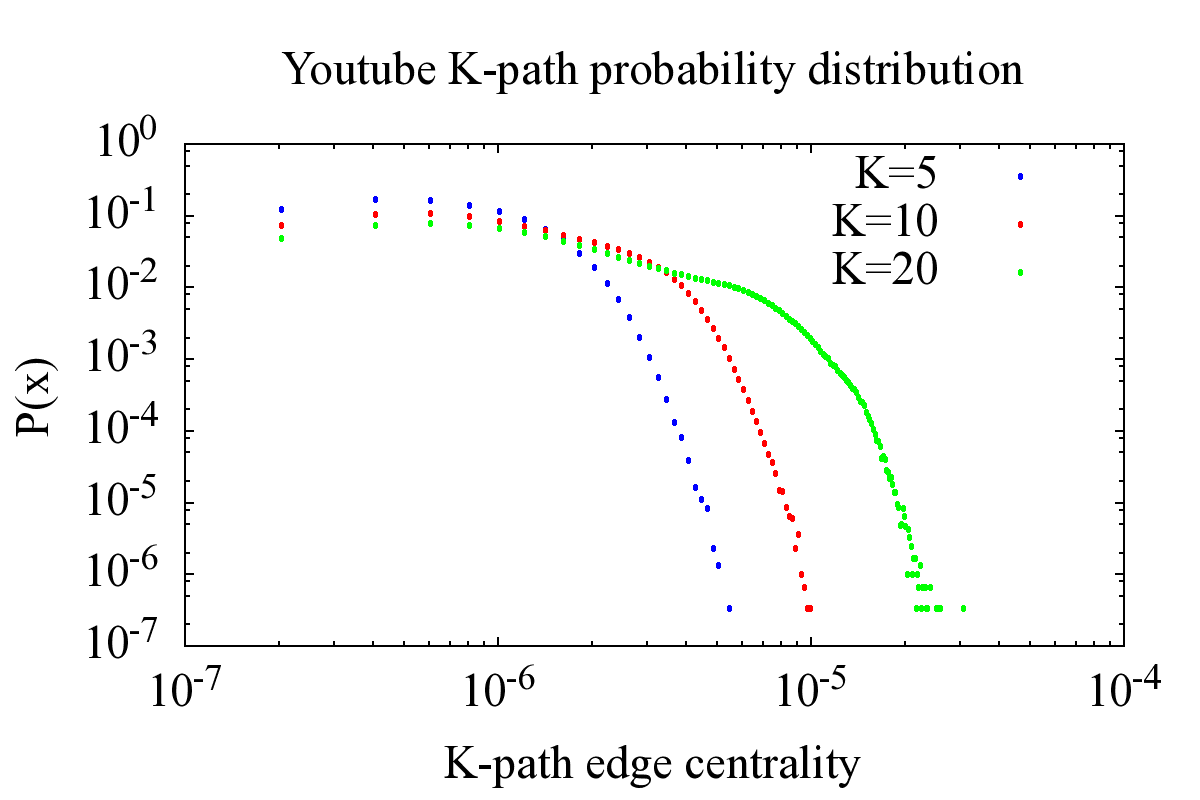}
	\caption{Effect of different $\kappa=5,10,20$ on networks described in Table \ref{tab:datasets}.}%
	\label{fig:k-correlation}%
\end{figure}

In the second experiment, we studied how the value of $\kappa$ impacted on edge centrality. In
detail, we considered the datasets separately and repeated the experiments described above. Also
for this experiment we considered three different values for $\kappa$, namely $\kappa = 5, 10, 20$. The
corresponding results are plotted in Figure \ref{fig:k-correlation}, where the probability $P$ of
finding an edge in the network which has the given value of centrality is plotted as a function of
the $\kappa$-path centrality. Each plot adopts a \emph{log-log} scale.

The analysis of this figure highlights three relevant facts:

\begin{itemize}
\item The probability of finding edges in the network with the lowest $\kappa$-path edge centrality values is smaller 			than finding edges with relatively higher centrality values.
			This means that the most of the edges are exploited for the message propagation by the random walks a number of 			times greater than zero.

\item The power law distribution in edge centrality emerges even more for different values of
 	    $\kappa$ and in presence of different datasets. In other words, if we use different values of
   	  $\kappa$ the centrality indexes may change (see below); however, as emerges from Figure
   \ref{fig:k-paths}, for each considered dataset, the curves representing $\kappa$ path
   centrality values are {\em straight} and {\em parallel} lines with the exception of the
   latest part. This implies that, for a fixed value of $\kappa$, say $\kappa = 5$, an edge
   $\overline{e}$ will have a particular centrality score. If $\kappa$ passes from 5 to 10 and,
   then, from 10 to 20, the centrality of $\overline{e}$ will be increased by a constant
   factor. This implies that the ordering of the edges remains unchanged and, therefore, the edge
   having the highest centrality at $\kappa = 5$ will continue to be the most central edges
   also when $\kappa = 10$ and $\kappa = 20$. This highlights a nice feature of WERW-Kpath: 
   potential uncertainties on the tuning of the parameter $\kappa$ do not have a
   devastating impact on the process of identifying the highest ranked edges.

\item The higher $\kappa$, the higher the value of centrality indexes. This has an intuitive
      explanation. If $\kappa$ increases, our algorithm manages longer paths to compute
      centrality values. Therefore, the chance that an edge is selected multiple times increases
      too. Each time an edge is selected, our algorithm awards it by a bonus score (equal to
      $\beta$). As a consequence, the larger $\kappa$, the higher the number of times an edge with high
      centrality will be selected, and ultimately, the higher its final centrality index.

      Such a consideration provides a practical criterion for tuning $\kappa$. In fact, if we
      select high values of $\kappa$, we are able to better discriminate edges with high
      centrality from edges with low centrality. By contrast, in presence of low values of
      $\kappa$, edge centrality indexes tend to edge flatten in a small interval and it is harder
      to distinguish high centrality edges from low centrality ones.

      On the one hand, therefore, it would be fine to fix $\kappa$ as high as possible.
      On the other, since the complexity of our algorithm is $O(\kappa m)$, large values
      of $\kappa$ negatively impact on the performance of our algorithm.
      A good trade-off (explained by the experiments showed in this section) is to fix $\kappa = 20$.
\end{itemize}

\section{Applications of our approach in Knowledge-Based systems}
\label{sec:applications}

In this section we detail some possible applications of our approach to rank edges in social
networks in the area of Knowledge-Based systems (hereafter, KBS).

In detail, we shall focus on three possible applications. The first is {\em data clustering} and we
will show how our approach can be employed in conjunction with a clustering algorithm with the aim
of better organizing data available in a KBS. The second is related to the {\em Semantic Web} and
we will show how our approach can be used to assess the strength of the semantic association
between two objects and how this feature is useful to improve the task of discovering new knowledge
in a KBS. The third, finally, is related to better understand the relationship and the roles of
user in virtual communities; in this case we show that our approach is useful to elucidate
relationships like trust ones.

\subsection{Data Clustering}
\label{sub:dataclustering}

A central theme in KBS-related research is the design and implementation of effective data
clustering algorithms \cite{SrMu94}. In fact, if a KBS has to manage massive datasets (potentially
split across multiple data sources), clustering algorithms can be used to organize available data
at different {\em levels of abstraction}. The end user (both a human user or a software program)
can focus only on the portion of data which are the most relevant to her/him rather than exploring
the whole data space managed by a KBS \cite{SrMu94,CaHaKh09,MaAb10}. If we ideally assume that any
data managed by a KBS is mapped onto a point of a multidimensional space, the task of clustering
available data requires to compute the mutual distance existing between any pair of data points.

Such a task, however, is in many cases {\em unfeasible}. In fact, the computation of the distance
can be prohibitively time-consuming if the number of data points is very large. In addition, KBS
often manage data which are related each other but, for these kind of data, the computation of a
distance could make no-sense: think, for instance, of data on health status of a person and her/his
demographic data like age or gender.

Therefore, many authors suggest to represent data as {\em graphs} such that each node represents a
data point and each edge specifies the type of relationships binding two nodes. The problem of
clustering graphs has been extensively studied in the past and several algorithms have been
proposed. In particular, the graph clustering problem in the social network literature is also
known as {\em community detection} problem \cite{fortunato2010community}.

One of the early algorithms to find communities in graphs/networks was proposed by Girvan and
Newman in 2002 \cite{girvan2002community}. Unfortunately, due to its high computational complexity,
the Girvan-Newman algorithm can not be applied on very large and complex data repositories
consisting of million of information objects.

Our algorithm, instead, can be employed to rank edges in networks and to find communities. This is
an ongoing research effort and the first results are quite encouraging \cite{isda2011a}.

Once a community finding algorithm is available we can design complex applications to effectively
manage data in a KBS. For instance, in \cite{xia2011community} the authors focused on online social
networks like Internet newsgroups and chat rooms. They analyzed through semantic tools the text
comments posted by users and this allowed large online social networks to be mapped onto weighted
graphs. The authors showed that the discovery of the latent communities is a useful way
to better understand patterns of interactions among users and how opinions spread in the network.

We then describe two use cases possibly benefiting from community detection algorithms.
In the first case, consider a social network in which users fill a profile specifying their
interests. A graph can be constructed which records users (mapped onto nodes) and relationship
among them (e.g., an edge between two nodes may indicate that two users share at least one
interest). Our algorithm, therefore, could identify group of users showing the same interests.

Therefore, given an arbitrary message (for instance a commercial advertisement) we could identify
groups of users interested to it and we could selectively send the message only to interested
groups.

As an opposite application, we can consider the objects generated within a social media platform.
These objects could be for instance photos in a platform like Flickr or musical tracks in a
platform like Last.fm. We can map the space of user generated contents onto a graph and apply on it
our community detection algorithm. In this way we could design advanced query tools: in fact, once a
user issues a query, a KBS may retrieve not only the objects exactly labeled by the keywords
composing user queries but also objects falling in the same community of the retrieved objects. In
this way, users could retrieve objects of their interest even if they are not aware about their
existence.

\subsection{Semantic Web}
\label{sub:semanticweb}

A further research scenario that can take advantage from our research work is represented by the
Semantic Web. In detail, Semantic Web tools like RDF allow complex and real-life scenarios to be
modeled by means of networks. In many cases these networks are called multi-relational networks (or
semantic networks) because they consist of heterogeneous objects and many type of relationships can
exist among them \cite{Rodriguez08}.

For instance, an RDF knowledge base in the e-learning domain  \cite{LaTs09} could consist of
students, instructors and learning materials in a University. In this case, the RDF knowledge base
could be converted to a semantic network in which nodes are the players described above. Of course,
an edge may link two students (for instance, if they are friends or if they are enrolled in the
same BsC programme), a student and a learning object (if a student is interested in that learning
object), an instructor and a learning material (if the instructor authored that learning material)
and so on \cite{Ursino-ODBASE03}.

A relevant theme in Semantic Web is to assess the {\em weight} of the relationships binding two
objects because this is beneficial to discover new knowledge. For instance, in the case of the e-learning
example described above, if a student has downloaded multiple learning objects on the same topic,
the weight of an edge linking the student and a learning material would reflect the relevance of
that learning material to the student. Therefore, learning materials can be ranked on the basis of
their relevance to the user and only the most relevant learning materials can be suggested to the
user.

An approach like ours, therefore, could have a relevant impact in this application scenario because
we could find interesting associations among items by automatically computing the weight of the
ties connecting them. To the best of our knowledge there are few works on the computation of node
centrality in semantic networks \cite{Rodriguez08} but, recently some authors suggest to extend
parameters introduced in Social Network Analysis like the concept of shortest path to
multi-relational networks \cite{RoWa10}.

Therefore, we plan to extend our approach to the context of semantic networks. Our aim is to use
simple random walks in place of shortest paths to efficiently discover relevant associations
between nodes in a semantic network and to experimentally compare the quality of the results
produced by our approach against that achieved by approaches relying on shortest paths.

\subsection{Understanding user relationships in virtual communities}
\label{sub:userunderstanding}

A central theme in KBS research is represented by the extraction of patterns of interactions among
humans in a virtual community and their analysis with the goal of understanding how humans
influence each other.

A relevant problem is represented by the classification of the relationship of humans on the basis
of their intensity. For instance, in \cite{Ding*11} the authors focus on the criminal justice
domain and, in particular, on the identification of social ties playing a crucial role in the transmission of
sensitive information. In \cite{Xia07}, the author provides a {\em belief propagation algorithm}
which exploits social ties among members of a criminal social network to identify criminals. Our
approach resembles that of \cite{Ding*11} because both of them are able too associate each edge in
a network with a score indicating the strength of the association between the nodes linked by that
edge.

A special case occurs when we assume that the edge connecting two nodes specifies a {\em trust
relationship} \cite{DeMeo*09,KiSo11}. In \cite{DeMeo*09}, the authors suggest to propagate trust
values along paths in the social network graph. In an analogous fashion, the approach of
\cite{KiSo11} uses path in the social network graph to propagate trust values and infer trust
relationships between pairs of unknown users. Finally, {\em Reinforcement Learning} techniques are
applied to estimate to what extent an inferred trust relationship has to be considered as credible.
Our approach is similar to those presented above because both of them rely on a {\em diffusion
model}. In \cite{DeMeo*09,KiSo11}, the main assumption is that trust reflects the {\em
transitive property}, i.e., if a user $x$ trusts a user $y$ who, in her/his turn, trusts a user
$z$, then we can assume that $x$ trusts $z$ too. In our approach, we exploit connections among
nodes to propagate messages by using simple random walks of bounded length. There are, however, some
relevant differences: in the approaches devoted to compute trust all the paths of any arbitrary length
are, in principle, useful to compute trust values even if the contribution brought in by long paths
is considered less relevant than that of short paths. Vice versa, in our approach, the length of a
path is bounded by a fixed constant $\kappa$.

\section{Conclusions} \label{sec:conclusions}

In this paper we introduced an edge centrality measure in social networks called $\kappa$-path edge centrality index.
Its computation is computationally feasible even on large scale networks by using the algorithm we provided.
It performs multiple random walks on the social network graph, which are simple and their length is bounded by a factor $\kappa$.
We showed that the worst-case time complexity of our algorithm is $O(\kappa m)$, being $m$ the number of edges in the social network graph.
Finally, we discussed experimental results obtained by applying our method to different online social network datasets.

We plan to extend our work in several directions. First of all, our centrality measure can be used
to detect communities in large social networks. Such a task is currently unfeasible if we use
classic measures like edge betweenness centrality. In fact, to the best of our knowledge, efficient
algorithms do not currently exist that estimate the community structure of a large network based on
global topological information and our strategy could fit well to this purpose. 
We believe that our approach could be beneficial in the field of visualization of large social networks as well.
In fact, recently it has been advanced the possibility of exploiting efficient network clustering techniques based on edge bundling to improve the graphical representation of the hierarchical structure of social networks \cite{jia2011social}.

In addition, we plan to design an algorithm to estimate the strength of ties between two social network actors: for instance, in social networks like Facebook this is equivalent to estimate the friendship degree between a pair of users. 

Finally, we point out that some researchers studied how to design parallel algorithms to compute centrality measures; for instance, \cite{madduri2009faster} proposed a fast and parallel algorithm to compute betweenness centrality. 
We guess that a new, interesting, research opportunity is to design parallel algorithms to compute the $\kappa$-path edge centrality.


\section*{Acknowledgments}
We would like to thank the Editor and the anonymous Referees whose comments helped us to greatly improve the quality of the work.


\section*{Appendix}

In this section we shall analyze the correctness of our ERW-KPath and WERW-KPath algorithms. In
details, we will study how the centrality indexes returned by these algorithms are related to the
actual centrality values provided in Definition \ref{def:k-path-edge-centrality}.

To explain our results it is convenient to re-write Equation (\ref{eq:edge-k-path}) in a more
manageable fashion. First of all let us consider an undirected graph $G = \langle V, E \rangle$ and
denote as $\phi_{sl}$ an arbitrary simple path in $G$ starting from a {\em fixed source node} $s$
and of length $l$ (i.e., the considered path contains $l$ edges). The graph $G$ can be {\em
unweighted} as well as {\em weighted}.

In the following, when it does not generate confusion, we will avoid subscripts to denote both
nodes and edges and, therefore, we will denote a node as $v$ (rather than $v_n$) and an edge as $e$
(rather than $e_m$).

Let us assume that the sequence of nodes forming $\phi_{sl}$ is $\phi_{sl} = \{s, u_1, \dots,
u_{l-1}\}$ (note that $s = u_0$); in addition, let us denote as $P(\phi_{sl})$ the probability of generating the path
$\phi_{sl}$ by simulating a simple path of length $l$. In \cite{alahakoon2011kpath-techrep} the
authors show that, in case $G$ is unweighted, the value of $P(\phi_{sl})$ is as follows

\begin{equation}
\label{eq:probunweighted}
P(\phi_{sl}) = \prod_{j = 1}^{l} \frac{1}{|O(u_{j-1}) - \{s, u_1, \dots, u_{j-2}\} |}
\end{equation}

Here $O(u_j)$ is the set of nodes adjacent to $u_j$ (i.e., a node $v$ belongs to $O(u_j)$ if there
is an edge joining $u_j$ to $v$).

In an analogous fashion, it is possible to consider the case of weighted graphs. In detail, let
$W(u,v)$ be the weight of the edge going from the node $u$ to the node $v$; in such a case, on the
wake of the considerations presented in \cite{alahakoon2011kpath-techrep}, we can derive the following
expression for $P(\phi_{sl})$

\begin{equation}
\label{eq:probweighted}
P(\phi_{sl}) = \prod_{j = 1}^{l}{\frac{W(u_{j-1},u_j)}{\sum_{v \in O(u_{j-1}) - \{s, \dots,u_{j-2}\}} W(u_{j-1},v)}}
\end{equation}

We are now able to re-write the expression of edge centrality index $L^{\kappa}(e)$ in terms of
$P(\phi_{sl})$. In detail, given an edge $e \in E$ and a path $\phi_{sl}$, we will use the notation
$e \in \phi_{sl}$ if the edge $e$ belongs to the path $\phi_{sl}$; we can therefore define a
variable $\chi(e \in \phi_{sl})$ as follows

$$
\chi(e \in \phi_{sl}) = \left\{
    \begin{array}{ll}
      1    & \mbox{if $e \in \phi_{sl}$} \\
      0  & \mbox{otherwise} \\
    \end{array}
  \right.
$$

Due to these definitions, it is possible to show that the edge centrality of an edge $e$ can be
rewritten as follows

\begin{equation}
\label{eqn:centrality}
L^{\kappa}(e) = \sum_{s \in V} \sum_{1 \leq l \leq \kappa} \sum_{\phi_{sl}} P(\phi_{sl}) \cdot \chi(e \in \phi_{sl})
\end{equation}

The interpretation of Equation (\ref{eqn:centrality}) is as follows. To compute the edge centrality
of an edge $e$ we start by fixing an arbitrary source node $s$. We consider a simple path
$\phi_{sl}$ starting from $s$ of length $l$. The path $\phi_{sl}$ contributes to the centrality of
$e$ only if it contains $e$ itself. This is captured by the product $\phi_{sl} \cdot \chi(e \in
\phi_{sl})$ in Equation (\ref{eqn:centrality}): in fact, if $e \in \phi_{sl}$, then $\chi(e \in
\phi_{sl}) = 1$ by definition and, therefore, the contribution of $\phi_{sl}$ to $L^k(e)$ is equal
to $P(\phi_{sl})$.

By contrast, if $e \notin \phi_{sl}$, then $\chi(e \in \phi_{sl}) = 0$ and, therefore, the path
$\phi_{sl}$ does not provide any contribution to the computation of $L^k(e)$.

Due to Definition \ref{def:k-path-edge-centrality}, in the computation of $L^k(e)$ we are
interested in {\em all} the simple paths up to length $\kappa$; this explains why, in Equation
(\ref{eqn:centrality}), we need a double sum over all the simple paths of length $l$ being $ 1 \leq
l \leq \kappa$.
Moreover, Definition \ref{def:k-path-edge-centrality} requires to consider all the nodes $s \in V$
as potential source nodes and this explains the third sum appearing in Equation
(\ref{eqn:centrality}).

It is also interesting to observe that the term $\sum_{1 \leq l \leq \kappa} \sum_{\phi_{sl}}
P(\phi_{sl}) \cdot \chi(e \in \phi_{sl})$ can be linked to the probability of selecting an edge $e
\in E$ under the assumption that a simple random path starts from a {\em fixed vertex} $v \in V$.
This is expressed by the following theorem:

\begin{theorem}
\label{th:selecting} Let $G = \langle V, E\rangle$ be a graph, $s \in V$ be
a node in $G$ and $e \in E$  be an edge in $G$. The probability $P_{e,s}$ of selecting
the edge $e$ by means of a simple random path starting from $s$ is $P_{e,s} = \sum_{1 \leq l \leq
\kappa} \sum_{\phi_{sl}} P(\phi_{sl}) \cdot \chi(e \in \phi_{sl})$.
\end{theorem}

To prove this result, let us focus on a vertex $s \in V$ and on an edge $e \in E$. We can consider
three cases:

{\em Case 1}. There is no simple path of length $ l \leq \kappa$ starting from $s$ and containing
$e$. In such a case, the term $\chi(e \in \phi_{sl})$ will be always 0 and, therefore, the value of
$P_{e,s}$ will be 0. Such a result is correct because, in this case, the probability of selecting
$e$ is clearly 0.

{\em Case 2}. There exists {\em exactly} one path $\phi_{sl}^{*}$ containing the edge $e$; if this
path is selected, then, the edge $e$ will be selected too and then the term $\chi(e \in \phi_{sl})$
will be equal to 1. The probability of selecting the edge $e$ will be, therefore, equal to the
probability of selecting the path $\phi_{sl}^{*}$ passing through $e$. In such a case the term
$P_{e,s}$ would simply be equal to $P_{e,s} = P(\phi_{sl}^{*})$ which coincides with the
probability of selecting $e$.

{\em Case 3}. There are multiple paths starting from $s$ and going through $e$. In such a case, the
probability of selecting $e$ is equal to the probability of selecting {\em at least} one of these
paths. Since the paths are generated one by one, the probability $P_{e,s}$ of selecting $e$ is
equal to $\sum_{1 \leq l \leq \kappa} \sum_{\phi_{sl}} P(\phi_{sl})$.

Once we provided a formal definition of edge centrality we are interested in analyzing the
centrality value generated by our algorithm. Let us focus on an edge $e$ and observe that our
algorithm performs $\rho$ trials and, in each trial, it generates a simple random path of at most
$\kappa$ edges. Let us consider the $i$-th trial and observe that the edge $e$ can be selected in
the $i$-th trial or not; of course, since the path must be simple, the edge $e$ can be selected no
more than once in a trial.
To model the selection of an edge $e$ in the generic, $i$-th trial, we define the random variable
$X_i(e)$ as follows

$$
X_i(e) = \left\{
    \begin{array}{ll}
      1  & \mbox{if $e$ has been selected in the $i$-th trial} \\
      0  & \mbox{otherwise}\\
    \end{array}
  \right.
$$

Recall that our ERW-KPath algorithm (along with its weighted version WERW-KPath) initially awards
any edge by assigning it a centrality index equal to $\frac{1}{|E|}$. Any time an edge $e$ is
selected, it gets an additional award equal to $\beta=\frac{1}{E}$; as a consequence, since the number of
times the edge $e$ is selected is equal to $\sum_{i = 1}^{\rho}X_i(e)$, the value $\omega(e)$
returned by the algorithm is equal to

\begin{equation}
\label{eqn:centralityapprox}
\omega(e) = \sum_{i=1}^{\rho} \frac{X_i(e) + 1}{|E|}
\end{equation}

Our goal is to show that the ERW-KPath and WERW-KPath algorithms provide a ``good''
approximation of $L^{\kappa}(e)$.
This is formalized by Theorem \ref{th:bounds}

\begin{theorem}
\label{th:bounds} Let $G = \langle V, E \rangle$ be a graph and, for each edge $e \in E$, let
$L^{\kappa}(e)$ be the $\kappa$-path edge centrality index of $e$ computed according to Definition
\ref{def:k-path-edge-centrality}. Finally, let $\rho$ be an integer. The following results hold
true:
\begin{enumerate}

\item The edge centrality value $\omega(e)$ computed by the ERW-Kpath algorithm on $G$ is
    related to the actual centrality value $L^{\kappa}(e)$ by the following relation:
    $\omega(e) = \frac{1}{|E|} + \frac{\rho}{|E||V|}L^{\kappa}(e)$

\item The edge centrality value $\omega(e)$ computed by the WERW-Kpath algorithm on $G$ is
    related to the actual centrality value $L^{\kappa}(e)$ by the following relation: $\xi^{'}
    L^{\kappa}(e) + \frac{1}{|E|} \leq \omega(e) \leq \xi^{''} L^{\kappa}(e) + \frac{1}{|E|}$,
    being $\xi^{'}$ and $\xi^{''}$ two suitable constants whose value is proportional to the
    ratio $\frac{\rho}{|E|}$.
\end{enumerate}

\end{theorem}

{\em Proof}. We shall consider two cases, depending on the fact that we decide to apply the ERW-Kpath or
WERW-Kpath algorithm.

{\em Case 1: ERW-Kpath}. Let us compute the expectation of both the members of Equation
(\ref{eqn:centralityapprox}). Due to the linearity of the expectation operator we get

$$
E\left[\omega(e)\right] = \sum_{i=1}^{\rho} \frac{E[X_i(e)]}{|E|} + \frac{1}{|E|}
$$

Observe now that, since $\omega(e)$ is a fixed value computed by our algorithm, then $E[\omega(e)]
= \omega(e)$.  As for $E[X_i(e)]$ it is simply equal to $P(X_i(e) = 1)$ due to the definition of
expectation
$$
E[X_i(e)] = 0 \cdot P(X_i(e) = 0) + 1 \cdot P(X_i(e) = 1) = P(X_i(e) = 1)
$$
Observe that $P(X_i(e) = 1)$ is the probability of selecting the edge $e$. Observe that the
ERW-Kpath algorithm manages an overall number of source nodes $s$ equal to $|V|$ and that each node
is selected uniformly at random. In addition, due to Theorem \ref{th:selecting}, once $s$ has been
fixed the probability of selecting $e$ starting from $s$ is equal to $\sum_{1 \leq l \leq \kappa}
\sum_{\phi_{sl}} P(\phi_{sl}) \cdot \chi(e \in \phi_{sl})$; the probability $P(\phi_{sl})$, in the
case of the ERW-Kpath algorithm, has to be intended as in Equation (\ref{eq:probunweighted}).

Due to these reasons, we get that

$$
P(X_i(e) = 1) = \frac{1}{|V|}\sum_{s \in V} \sum_{1 \leq l \leq \kappa} \sum_{\phi_{sl}}
P(\phi_{sl}) \cdot \chi(e \in \phi_{sl})
$$

which can be rewritten as

$$
P(X_i(e) = 1) = \frac{1}{|V|}L^{\kappa}(e)
$$

Due to this result we can write
$$
\omega(e) =  \frac{1}{|E|} + \frac{1}{|E|}\sum_{i=1}^{\rho} \frac{1}{|V|}L^{\kappa}(e)
$$

After some simplifications we get

$$
\omega(e) = \frac{1}{|E|} + \frac{\rho}{|E||V|}L^{\kappa}(e)
$$

which states that the actual value of $L^{\kappa}(e)$ differs from that computed by our algorithm
by a constant factor.

{\em Case 2: WERW-Kpath}. The proof in this case is analogous to Case 1. In detail, by repeating
the considerations provided in Case 1, we can show that

$$
\omega(e) = \sum_{i=1}^{\rho} \frac{P(X_i(e) = 1)}{|E|} + \frac{1}{|E|}
$$

In such a case, however, the expression for $P(X_i(e) = 1)$ is slightly more complex than in Case
1. In detail, in the WERW-Kpath algorithm the source node $s$ is selected with probability $P(s)$
provided in Equation (\ref{eq:p-e-n}). Therefore, we get

$$
P(X_i(e) = 1) = \sum_{s \in V} P(s) \sum_{1 \leq l \leq \kappa} \sum_{\phi_{sl}}
P(\phi_{sl}) \cdot \chi(e \in \phi_{sl})
$$

and the term $P(\phi_{sl})$ is now computed according to Equation (\ref{eq:probweighted}) because
the graph $G$ is now weighted\footnote{Recall that the WERW-Kpath algorithm iteratively computes
the weight of each edge and the weight at the $n$-th iteration is proportional to $\omega_{n-1}$.}.
Set $\overline{P} = \max_{s \in V} P(s)$ and $\overline{p} = \min_{s \in V} P(s)$; we get the
following bounds

$$
\overline{p} \sum_{s \in V} \sum_{1 \leq l \leq \kappa} \sum_{\phi_{sl}}
P(\phi_{sl}) \cdot \chi(e \in \phi_{sl}) \leq P(X_i(e) = 1) \leq \overline{P} \sum_{s \in V} \sum_{1 \leq l \leq \kappa} \sum_{\phi_{sl}}
P(\phi_{sl}) \cdot \chi(e \in \phi_{sl})
$$

The last equation can be rewritten as

$$
\overline{p} L^{\kappa}(e) \leq P(X_i(e) = 1) \leq \overline{P} L^{\kappa}(e)
$$

and, by summing over all the indexes $i = 1 \dots \rho$

$$
\rho \overline{p} L^{\kappa}(e) \leq \sum_{i = 1}^{\rho}P(X_i(e) = 1) \leq \rho \overline{P} L^{\kappa}(e)
$$

and

$$
\frac{\rho \overline{p} L^{\kappa}(e)}{|E|} + \frac{1}{|E|} \leq \frac{1}{|E|}\sum_{i = 1}^{\rho} P(X_i(e) = 1) + \frac{1}{|E|} \leq \frac{\rho \overline{P} L^{\kappa}(e)}{|E|} + \frac{1}{|E|}
$$

By setting $\xi^{'} = \frac{\rho\overline{p}}{|E|}$ and $\xi^{''} = \frac{\rho\overline{P}}{|E|}$
and by Equation (\ref{eqn:centralityapprox}), we obtain

$$
\xi^{'}L^{\kappa}(e) + \frac{1}{|E|} \leq \omega(e) \leq \xi^{''} L^{\kappa}(e) + \frac{1}{|E|}
$$

which ends the proof.




\bibliographystyle{elsarticle-num}
\bibliography{k-path-bib}


\end{document}